**Temperature dependence of the Casimir pressure on two graphene sheets placed on a substrate**

M.V. Davidovich
Saratov National Research State University named after N.G. Chernyshevsky, 410012 Saratov, Russia
E-mail: davidovichmv@yamdex.ru

The dispersion interaction between two infinite sheets of graphene located on a substrate is considered based on the Lifshitz formula, taking temperature into account. The result is obtained in the form of a series that is easy to calculate. The influence of the interband conductivity of graphene, which turned out to be small, is examined. Formulas for the temperature correction of Casimir pressure are presented, based on the temperature dependence of graphene conductivity and the integral in the van Kampen–Schram form for the Lifshitz formula. In the presence of a non-dispersive substrate, the compressive pressure on graphene sheets is lower than for a vacuum gap.

## Introduction

The Casimir pressure $P$ on material plates arises due to the excess fluctuation energy density of the quantum vacuum outside the plates compared to the energy density inside (between the plates) [1,2]. This leads to the attraction of the plates, i.e., to negative pressure between the plates or to positive external pressure. The field does not penetrate perfectly conducting plates, and they can be considered as ideal screens with infinite conductivity. Therefore, the model for perfectly conducting, infinitely thin screens yields an interaction with the Casimir force [1]. In real structures, the Casimir force is primarily due to vacuum fluctuations associated with changes in boundary conditions [1], whereas thermal fluctuations usually make a significantly smaller contribution to it [2]. This contribution is significant in the far field, where the law of force action changes. Extended objects (dielectric and metal plates) are described by the dielectric permittivity (DP), the dependence of which on temperature is usually weak for dielectrics. For metals, this dependence is stronger, mainly due to changes in the collision frequency (CF). Typically, to analyze Casimir forces, a plasma model of Drude metals without collisions is used, which in pure metals corresponds well to reality only at zero temperature. The change in density with temperature for metals, i.e., the plasma frequency (PF), also has a slight effect on the Drude model. However, taking into account the CF in the Drude model significantly alters the interaction [3] and leads to a violation of Nernst's heat theorem [4–7].

Graphene is one of several well-known 2D materials that possess unique electrical, thermal, and mechanical properties. It has a thickness of one atomic layer and is described by two-dimensional conductivity $\sigma$, which is highly dependent on temperature $T$ and chemical potential $\mu_c$. Thermal correlations for graphene are related not to DP but to surface conductivity, i.e., they are determined not by Re($\varepsilon$) but by Im($\sigma$). Pure graphene at zero temperature has no conductivity: there are no electrons in its conduction band (CB), and there are no holes in the valence band (VB). However, it is still not possible to obtain graphene with an impurity concentration below $10^9$ $m^{-2}$. Impurities change the chemical potential of graphene $\mu_c$, and electrons appear in the CB, while holes appear in the VB. The chemical potential can also be changed by applying an electric potential. Temperature changes the Fermi-Dirac (FDF) function $f_{FD} = f^{\pm}$ in the subzones $\pi^{\pm}$ of the Brillouin zone (BZ)of graphene, which leads to the appearance of electrons in the CB and holes in the VB. Graphene interacts with a thermal field like with a Langevin thermostat, therefore it is described by a density matrix satisfying the Neumann equation [8], a quantum analogue of the Liouville equation. We consider graphene to be in equilibrium and described by a diagonal density matrix with components $f^+$ for the valence band and $f^-$ for the conduction band, where the sign corresponds to the sign of the energy $\pm\varepsilon$: $f^{\pm} = f_{FD}(\pm\varepsilon)$. In pure graphene, the concentration of charge carriers due to temperature is determined by the Fermi velocity $v_F$ and is equal to $n_e = n_h = \pi(k_B T)^2/(\hbar v_F)^2$. Here, the model of massless Dirac fermions is used.

The conductivity of graphene depends in a complex way on the chemical potential and temperature [9, 10]. For the first time, the band structure of graphene was obtained in [11]. The most accurate formula for its conductivity, taking into account spatial dispersion (SD) in a linear approximation based on the Kubo approach and the temperature Green Function method, was obtained in [12] in the form $\sigma_{\alpha\beta}(\omega,\mathbf{k}) = i\sigma_0\left(\sigma_{\alpha\beta}^{\text{intra}}(\omega,\mathbf{k}) + \sigma_{\alpha\beta}^{\text{inter}}(\omega,\mathbf{k})\right)$ where

$$\sigma_{\alpha\beta}^{\text{intra}}(\omega,\mathbf{k}) = \frac{4}{\pi^2}\sum_{l=1,2}\iint_{BZ}\frac{v^{\alpha}v^{\beta}\left[f_{FD}(\varepsilon_l(\mathbf{p}_-)) - f_{FD}(\varepsilon_l(\mathbf{p}_+))\right]d^2p}{\left[\varepsilon_l(\mathbf{p}_+) - \varepsilon_l(\mathbf{p}_-)\right]\left[\omega - \varepsilon_l(\mathbf{p}_+) + \varepsilon_l(\mathbf{p}_-)\right]}, \tag{1}$$

$$\sigma_{\alpha\beta}^{\text{inter}}(\omega,\mathbf{k}) = \frac{8\omega}{\pi^2}\iint_{BZ}\frac{v_{12}^{\alpha}v_{21}^{\beta}\left[f_{FD}(\varepsilon_1(\mathbf{p}_-)) - f_{FD}(\varepsilon_2(\mathbf{p}_+))\right]d^2p}{\left[\varepsilon_2(\mathbf{p}_+) - \varepsilon_1(\mathbf{p}_-)\right]\left\{\omega^2 - \left[\varepsilon_2(\mathbf{p}_+) - \varepsilon_1(\mathbf{p}_-)\right]^2\right\}}. \tag{2}$$

Here $l = 1,2$ corresponds to the conduction band of electrons and the valence band, respectively, i.e. $\varepsilon_1(\mathbf{p}) = \varepsilon^+(\mathbf{p})$, $\varepsilon_2(\mathbf{p}) = \varepsilon^-(\mathbf{p})$, $\varepsilon_l(\mathbf{p}_{\pm}) = \varepsilon_l(\mathbf{p}) \pm \mathbf{p}_k\omega/2$, where $\mathbf{p}_k = \hbar\mathbf{k}$ is the photon momentum, $\mathbf{p} = \hbar\mathbf{q}$ is quasi-momentum, $v^{\alpha}$ is the matrix element of the velocity operator of charge carriers (electrons and holes) between states in one zone, $v_{12}^{\alpha}$ is the matrix element of the velocity operator of carriers between states in different zones, $\alpha,\beta$ = $x,y$. The integration goes over the entire hexagonal BZ. However, in the approximation of massless Dirac fermions with dispersion $\varepsilon^{\pm}(\mathbf{k}) = \pm\hbar v_F q$, the conductivity is scalar and is determined by formulas [12–15].

$$\sigma^{\text{intra}}(\omega,\omega_c,\mu_c,T) = \frac{ie^2}{\pi\hbar^2(\omega - i\omega_c)}\int_0^{\infty}\left(\partial_{\varepsilon} f_{FD}(\varepsilon) - \partial_{\varepsilon} f_{FD}(-\varepsilon)\right)\varepsilon d\varepsilon, \tag{3}$$

$$\begin{aligned}\sigma^{\text{inter}}(\omega,\omega_c,q,\mu_c,T) = \frac{-ie^2(\omega - i\omega_c)}{\pi\hbar^2}\int_0^{\infty}\Bigg(&\frac{f_{FD}(-\varepsilon + q\hbar v_F/2)}{(\omega - i\omega_c - qv_F/2)^2 - 4\varepsilon^2/\hbar^2} - \\ &- \frac{f_{FD}(\varepsilon - q\hbar v_F/2)}{(\omega - i\omega_c + qv_F/2)^2 - 4\varepsilon^2/\hbar^2}\Bigg)d\varepsilon\end{aligned}, \tag{4}$$

in which $f_{FD}(\varepsilon) = \left[\exp((\varepsilon - \mu_c)/(k_B T)) + 1\right]^{-1}$ is the equilibrium FDF, $\mu_c$ is chemical potential. Conductivity (4) depends on the transverse wave number $q = \kappa = \sqrt{k_x^2 + k_y^2}$. Here, the integration is carried out over the infinite Dirac cones, which is a strong approximation. In works [13–15], these integrals are given for $\kappa = 0$, i.e., without taking into account the SD. Integral (3) at temperatures around room temperature and lower, and for $\mu_c < 1$ eV, is calculated with good accuracy in the form [12,15] (the dependence $\exp(i\omega t)$ is used here)

$$\sigma^{\text{intra}}(\omega,\omega_c,\mu_c,T) = \frac{k_B T(e^2/\hbar)\varphi(\mu_c,T)}{\pi\hbar\omega_c(1 + i\omega/\omega_c)} = \frac{4\sigma_0 k_B T\varphi(\mu_c,T)}{\pi\hbar\omega_c(1 + i\omega/\omega_c)}, \tag{5}$$

$$\varphi(\mu_c,T) = \ln\left(2\cosh\left(\frac{\mu_c}{2k_B T}\right)\right). \tag{6}$$

As for integral (4), it was calculated in work [15] for the case when $k_B T << \mu_c$, $k_B T << \hbar\omega$, $\kappa = 0$:

$$\sigma^{\text{inter}}(\omega,\omega_c,\mu_c,T) \approx \frac{-i\sigma_0}{\pi}\ln\left(\frac{2|\mu_c| - (\omega - i\omega_c)\hbar}{2|\mu_c| + (\omega - i\omega_c)\hbar}\right). \tag{7}$$

In this case, in integral (4) we can replace all the functions $f_{FD}$ with ones at the lower limit $|\mu_c|$ of the integral, from which we obtain formula (7). For $k_B T = 0$ and $\mu_c = 0$, the formula (7) implies $\sigma^{\text{inter}} = \sigma_0 = e^2/(4\hbar)$, and from (3) we have $\sigma^{\text{intra}} = 0$. The dimensionless graphene conductivity quantum $\xi_0 = \sigma_0\eta_0 = \pi\alpha_0 = 0.0229$ can be expressed in terms of the fine-structure constant $\alpha_0 = 1/137$. The above formulas are valid up to frequencies satisfying the condition $\hbar\omega < 3$ eV. In the UV range, photoionization of carbon atoms occurs first, with the breaking of $\pi$-bonds, followed by the breaking of $\sigma$-bonds; in this case, a plasma quantum model of a 2D electron gas should be used to describe the conductivity [16].

Formula (7) is not accurate even at room temperature, under conditions of low electrochemical potentials and low frequencies, even when $\kappa$=0. The most effective method is to numerically calculate integral (4). A more rigorous approach involves using model [12] and equations (1), (2) with integration over the entire Brillouin zone. Since dispersion interaction is determined by complex integrals, obtaining conductivity numerically is inconvenient, and it is desirable to have analytical expressions that take into account SD, which is one of the goals of this work. The results are conveniently presented in terms of the dimensionless chemical potential $\alpha = \mu_c / (k_B T)$ and the dimensionless complex frequency $\beta = \hbar(\omega - i\omega_c)/(2k_B T)$. Integral (4) depends strongly on the ratio of these quantities. In [14], the form of the tensor conductivity of graphene, taking into account the SD for intraband conductivity at zero temperature, was obtained in the first approximation:

$$\sigma_{xx}(\omega,\omega_c,\mathbf{k}) = \sigma^{\text{intra}}\left[1 + \frac{v_F^2}{4(\omega - i\omega_c)^2}\left(3 - \frac{2i}{\omega/\omega_c}\right)k_x^2 + \frac{v_F^2}{4(\omega - i\omega_c)^2}k_y^2\right] + \sigma^{\text{inter}}(k),$$

$$\sigma_{xy}(\omega,\omega_c,\mathbf{k}) = \sigma_{\text{intra}}\left[\frac{v_F^2}{2(\omega - i\omega_c)^2}k_x k_y\right], \tag{8}$$

$$\sigma_{yy}(\omega,\omega_c,\mathbf{k}) = \sigma^{\text{intra}}\left[1 + \frac{v_F^2}{4(\omega - i\omega_c)^2}\left(3 - \frac{2i}{\omega/\omega_c}\right)k_y^2 + \frac{v_F^2}{4(\omega - i\omega_c)^2}k_x^2\right] + \sigma^{\text{inter}}(k).$$

This result was obtained by solving the kinetic equation in the drift approximation and in the Bhatnagar–Gross–Krook approximation using an expansion in terms of a small parameter, and it is valid for small $\kappa$. However, in this case, the contribution of the SD to the dispersive force is small. Neglecting the Fermi velocity in relation to the speed of light $c$ in (4), we obtain the scalar interband conductivity without taking the SD into account. It should be noted that the integration over the infinite Dirac cones in (3) and (4) is a rather crude approximation. At a quantum energy of 1 eV or higher, the cones become curved and transition into a zone with a finite tensor effective mass of the particles [10]. The effective mass is zero only at the Dirac points themselves, and already in their vicinity it begins to differ from zero and becomes tensorial. The applicability of the given formulas is determined by the fact that the main contribution to the dispersion forces comes from low frequencies, where these effects are small. At low frequencies, the contribution (7) is small, and it is possible to limit oneself to intraband conductivity. However, this requires numerical justification, which is provided below. If interband conductivity is taken into account, its temperature-dependent form should be derived, which is the subject of the next section.

The temperature dependence of dispersion forces for graphene at low temperatures may manifest as a temperature dependence of conductivity when using interaction formulas that are independent of temperature. It is more rigorous to use temperature-dependent formulas. Since graphene sheets are constrained in two directions, the surface currents that arise in them have normal components that vanish at the boundaries. This means that the specific force for finite samples is less than the theoretical value for infinite samples. The surface conductivity of graphene is usually described by the Drude model $\sigma(\omega)=\sigma(0)/(1+i\omega/\omega_c)$. On a finite sheet with dimensions *Lx*=*Ly*=*L*, current resonances occur with the lowest resonant frequency $\omega_0 = c\sqrt{2}\pi/L$. This indicates the presence of an electron–atom bond in such a cluster: the electron cannot escape to infinity. Drude's formula is modified as follows:

$$\sigma(\omega)=\frac{\sigma(0)}{1+i\omega/\omega_c - i\omega_0^2/(\omega\omega_c)} \ . \tag{9}$$

This can be considered as the Drude–Smith model [17]. At high frequencies, it does not differ from the Drude model. At very low frequencies, the conductivity may become capacitive and small. Another aspect is related to CF. For pure graphene, it should depend on temperature according to the law $\omega_c(T)=\omega_c(T_0)T/T_0$ [18], where zero denotes room temperature. At high frequencies, the Drude formula gives a decrease in conductivity according to the laws $\sigma'(\omega)=\sigma(0)\omega_c^2/\omega^2$, $\sigma''(\omega)=-\sigma(0)\omega_c/\omega$, for the real and imaginary parts. Considering only contribution (5), we find $\sigma^{\text{intra}}(0,\omega_c,\mu_c,0)=e^2\mu_c/(2\pi\hbar^2\omega_c)=2\sigma_0\mu_c/(\pi\hbar\omega_c)$. This actual conductivity is related to the concentration of carriers, which is determined by the chemical potential. By determining the frequency $\Omega=2\mu_c/(\pi\hbar)$, we see that the conductivity at that frequency is equal to $\sigma_0$ and is small, and the dimensionless conductivity $\xi_0=0.0229$ is also small. It is convenient to represent

$$\sigma^{\text{intra}}(\omega,\omega_c,\mu_c,T)=\frac{\sigma(0,T)\Omega}{\omega_c+i\omega},$$

where $\sigma(0,T)=(e^2/(2\hbar\mu_c))k_BT\varphi(\mu_c,T)$. With such conductivity, the reflection coefficient (RC) at *T*=0 is of the order of 0.04 at the frequency $\Omega$, i.e., it can be quite reasonably used $10\Omega$ as a upper value. Taking the upper frequency limit $10\Omega$, we will obtain an RC of the order of 0.004, i.e., such high frequencies practically do not affect the force. At distances of the order of 1 nm and less, graphene sheets interact according to the van der Waals model. 1. If we take zero distance in the formula for the dispersion force, it will diverge, which is a consequence of the failure of the argument principle [19–22]. Small distances correspond to small wavelengths of the order of atomic ones. At the corresponding frequencies, the formulas for graphene conductivity are no longer valid, and graphene should be considered as a two-dimensional electron gas with ionized $\pi$- and $\sigma$-electrons. At small field periods, electrons oscillate at small distances near an atom, practically without scattering. This can be interpreted as a decrease in CF with an increase in frequency. If you use the model

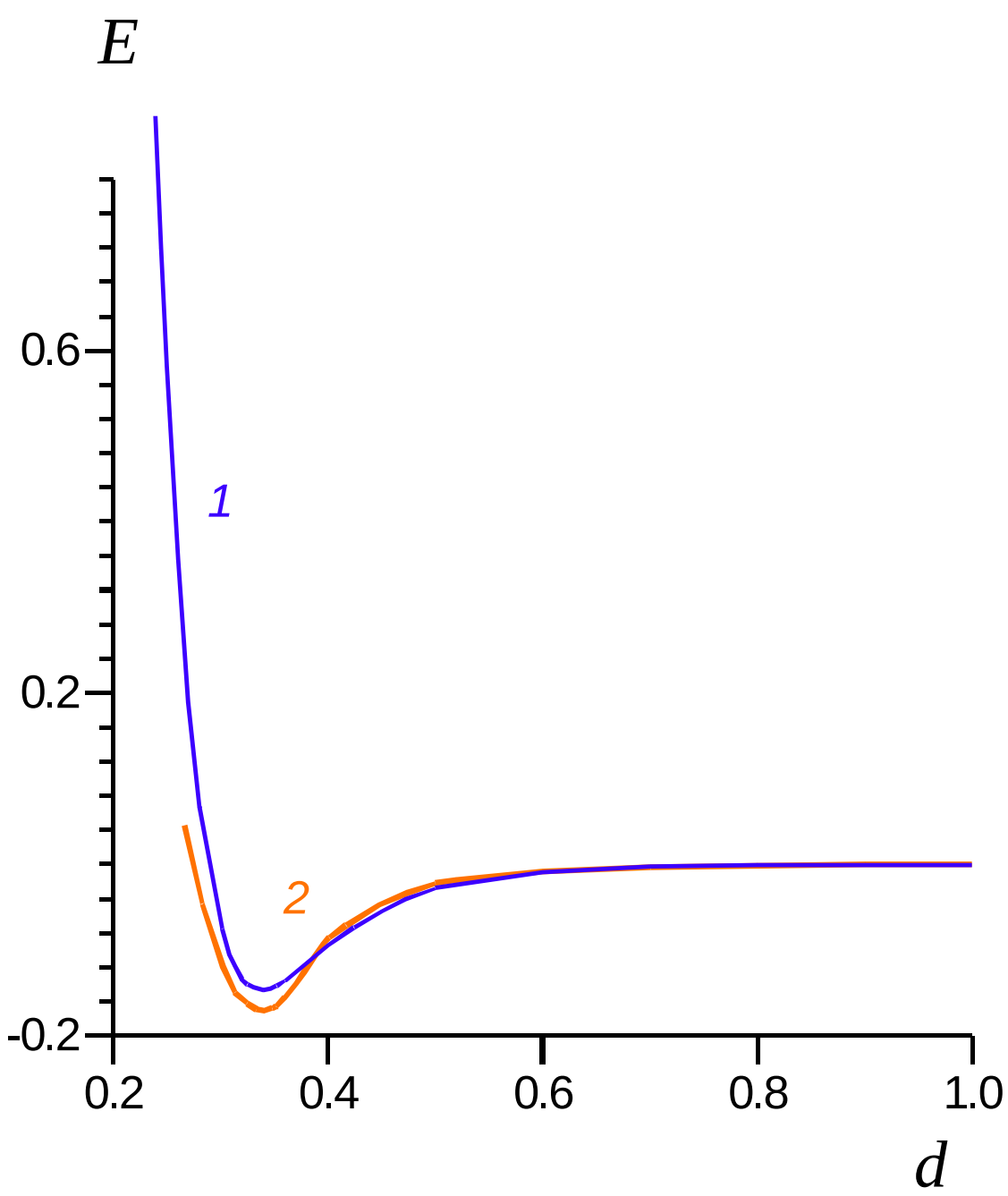

Fig. 1. Total binding energy (eV) per hexagon (curve 1) as a function of distance *d* (nm), calculated using the DFT method with consideration of van der Waals interaction using the Lennard-Jones potential (curve 1) and the 3D Grimme method (2).

$$\omega_c(\omega,T)=\frac{\omega_{c0}(T)}{1+(\omega/\Omega)^{\nu}}, \qquad (10)$$

then, without using a frequency restriction, it is possible to achieve the convergence of the integrals in all cases, including *d*=0. In (10) $\omega_{c0}(T)$ is the low-frequency CF depends on temperature. In fact, this means that the conductivity decreases more rapidly with frequency. It is sufficient to take $\nu \geq 3$. The calculations show a weak dependence of the result on $\nu$, since at low frequencies the conductivity depends weakly on $\nu$.

*The aim of the work* is to obtain the dependence of the external Casimir pressure on two graphene sheets in a vacuum and two graphene sheets on a substrate at *T*=0 and an arbitrary temperature. This pressure at zero temperature is determined by formula [19–23]

$$P(d)=-\frac{\hbar c}{2\pi^2}\int_0^\infty\int_0^\infty K(k,\kappa)\left(\frac{1}{f_e(\kappa,k,d)}+\frac{1}{f_h(\kappa,k,d)}\right)\kappa d\kappa dk\,, \qquad (11)$$

and at high temperatures, the formula [24.25] applies

$$P(d,T)=-\frac{k_BT}{\pi}\sum_{n=0}^{\infty}\frac{k_n^3}{1+\delta_{n0}}\int_1^\infty\sum_{\alpha=e,h}\frac{\varepsilon^{3/2}(k_n,p)p^2dp}{r_{n\alpha}^{-2}\exp\left(2pk_n\varepsilon^{1/2}(k_n,p)d\right)-1}. \qquad (12)$$

In (11) $k=ik_0=i\omega/c$, $K(k,\kappa)=\sqrt{\kappa^2+\varepsilon(k)k^2}$, $f_{e,h}(\kappa,k,d)$ are the dispersion equations (DE) for E- and H-waves in a multilayer structure, generally speaking. In [23–25], a vacuum gap between infinite dielectric plates is considered: $r_e=(s-\varepsilon p)/(s+\varepsilon p)$, $r_h=(s-p)/(s+p)$, $s=\sqrt{p^2-1+\varepsilon}$, $\xi=i\omega$. For two graphene sheets on a dielectric substrate with a DP $\tilde{\varepsilon}(\omega)$, we have the DE $f_{e,h}(\kappa,k,d)=r_{e.h}^{-2}\exp(2Kd)-1$. Here $r_{n\alpha}^{-2}$ are the inverse squares of the reflection coefficient (RC) from the structure with graphene on the vacuum side or on the substrate side for E-modes and H-modes, and *d* is the thickness of the substrate or the vacuum gap. We assume that for an ideal dielectric with a dielectric permittivity $\tilde{\varepsilon}$ that is independent of frequency within the considered frequency spectrum, the permittivity can be taken out from under the integral. In the case of a metallic substrate, we take dispersion into account and denote the permittivity as $\varepsilon(\omega)$ or $\varepsilon(k)$. The considered configurations are shown in the inserts, Fig. 2. The integration in (11) is carried out over $k$ and over $\kappa=\sqrt{k_x^2+k_y^2}$. In formula (12) $k_n=2\pi k_BTn/(\hbar c)=2\pi n/\delta_T$, where $\delta_T=\hbar c/(k_BT)$ is the characteristic thermal length associated with temperature, and $\delta_{n0}$ is the Kronecker symbol. Formula (11) is transformed by substituting $\kappa^2=k_x^2+k_y^2=k^2\varepsilon(k)(p^2-1)$, $K=\sqrt{k^2\varepsilon(k)+\kappa^2}=k\sqrt{\varepsilon(k)}p$, $\kappa d\kappa=k^2\varepsilon(k)pdp$ into the form $P(d)=P_e(d)+P_h(d)$, where

$$P_{e,h}(d)=-\frac{\hbar c}{2\pi^2}\int_1^\infty dpp^2\int_0^\infty\frac{k^3\varepsilon^{3/2}(k)}{r_{e,h}^{-2}(p,k)\exp\left(2pk\sqrt{\varepsilon(k)}d\right)-1}dk\,, \qquad (13)$$

and then by replacing $y=2pk\sqrt{\varepsilon(k)}d$ to the form

$$P_{e,h}(d)=-\frac{\hbar c}{32\pi^2d^4}\int_1^\infty\frac{dp}{p^2}\int_0^\infty\frac{\varepsilon^{-1/2}(p,y)y^3}{r_{e,h}^{-2}(p,y)\exp(y)-1}dy\,. \qquad (14)$$

Formula (13) takes into account only the zero fluctuations $\hbar\omega/2$ of the quantum vacuum. Adding thermal fluctuations to them by introducing a factor $\coth(\hbar kc/(k_BT))$ and switching from integration over *k* to summation over Matsubara frequencies or wave numbers $k_n$ leads, for high temperatures, to formula (12). Formally, this formula is obtained by Wick rotation [2]. However, Lifshitz did not take into account the contour integral [26,27].

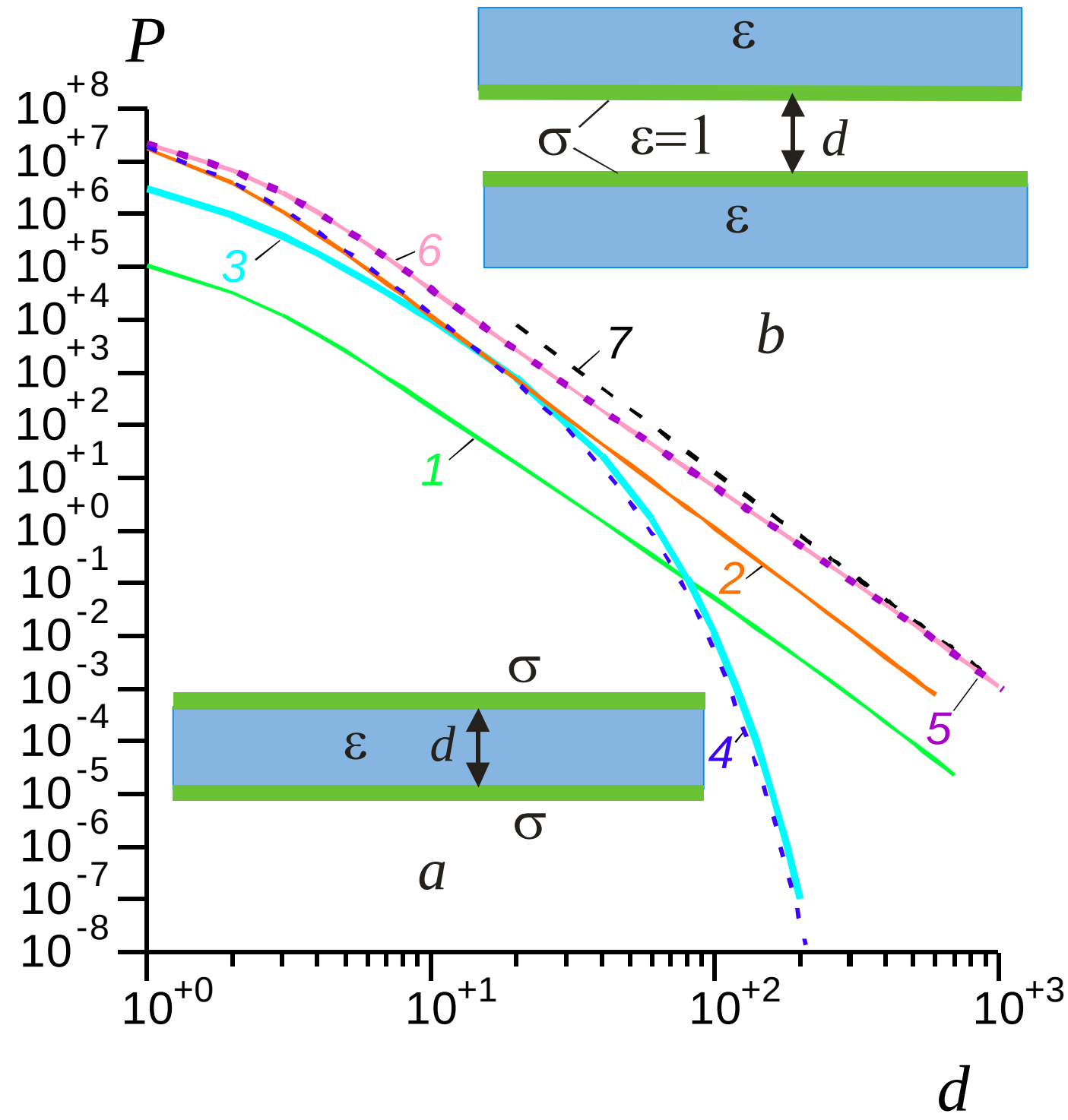

Fig. 2. External pressure $P$ (N/m$^2$) as a function of thickness $d$ (nm) for: graphene sheets in vacuum (curve 1); graphene sheets on a dielectric substrate $\varepsilon$=10 (2); graphene sheets between the plasma layer $\varepsilon_L$=1, $\omega_p$=1.6×10$^{16}$ Hz (3), $\omega_c$=0; graphene sheets on a metal substrate $\varepsilon_L$=10, $\omega_p$=1.6×10$^{16}$ Hz, $\omega_c$=0 (4); thick metal plates separated by a vacuum gap $d$ (5); thick metal plates with graphene sheets separated by a vacuum gap $d$ (6). For curves 5 and 6 $\omega_c$=3.2×10$^{13}$ Hz. $T$=300 K. Casimir's result is shown by line 7.

It is small for high temperatures and can be neglected [26]. However, at extremely low temperatures, it makes a significant contribution, and formula (12) yields a significantly smaller contribution than (11) [26].

Some remarks should be made regarding formula (11). It was first presented in Lifshitz's works [23–25] and has been proven in a number of works [19–22, 28–30] using several different methods, including the variational method [28], the principle of argument [19–22], the method of quantum statistical field theory [29], and the application of the theory of equilibrium fluctuations [30]. At the same time, formulas (11) and (13) imply the result of Casimir [1] if $r_{e,h}^2 = 1$. This result corresponds to the far zone ($d \to \infty$), as does Casimir's result itself (in the Euler–Maclaurin formula, Casimir implicitly assumed $d \to \infty$). In the near field, Casimir's force and energy must be finite and are determined by the van der Waals interaction. In [31], Casimir's method for calculating the free energy of a cavity with ideal walls at finite temperature is applied, and corrections to [1] are obtained. At the same time, the result for $d$~0 does not correspond to the result (12) [26]. We will use (12) at high temperatures. At low temperatures, we will use (11) with the temperature dependence of the conductivity.

Formula (11) follows most simply from the "argument principle" theorem [19–21]. Specifically, if $\tilde{\omega}_n(\kappa)$ are the complex roots (dispersion branches) of the equations $f_{e,h}(\omega,\kappa)=0$, then their sum is determined by an integral in the complex plane:

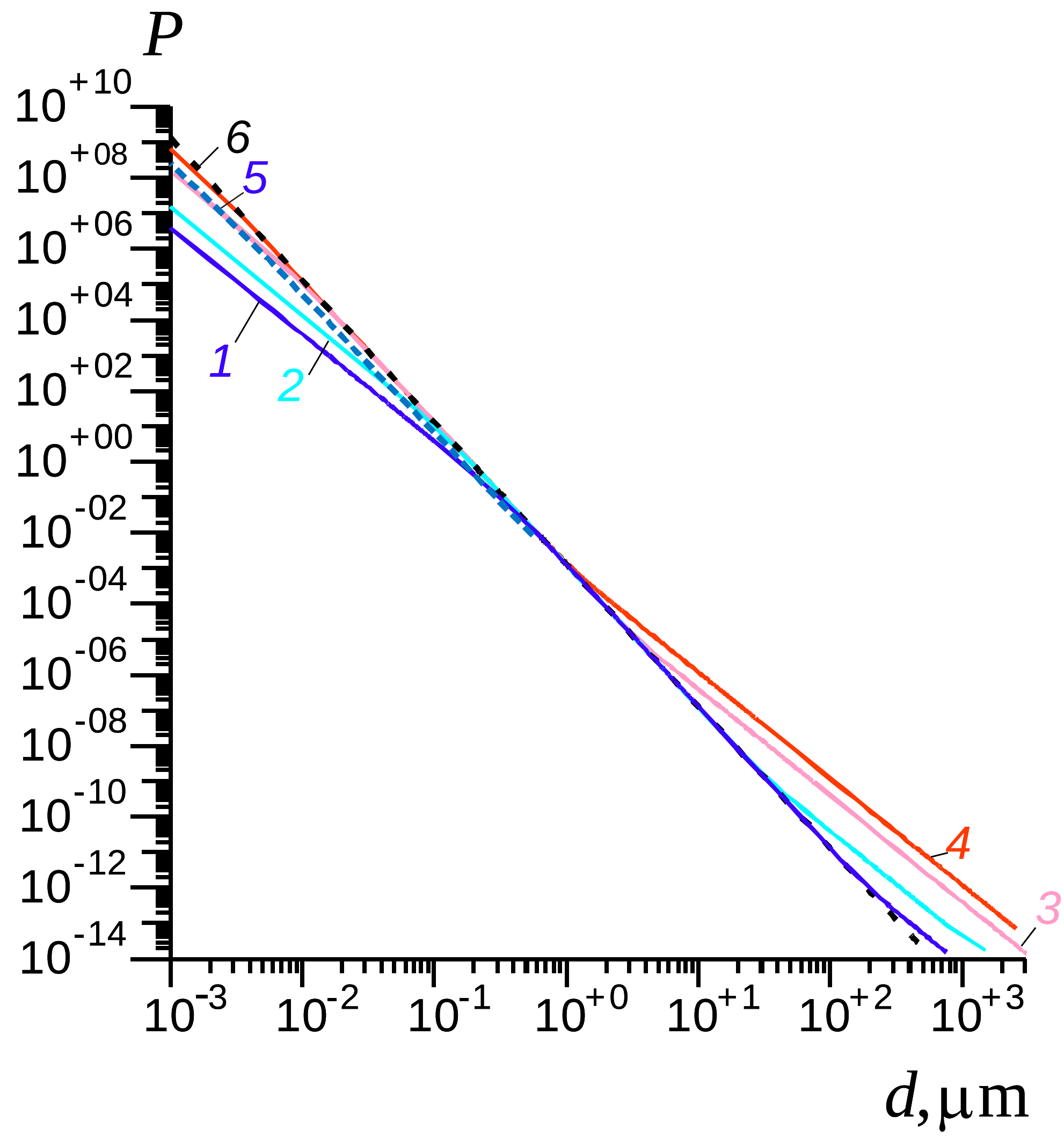

Fig. 3. External Casimir pressure $P$ (N/m$^2$) on two ideally conductive metal layers depending on $a$ (in μm) at different temperatures (formula (6)): 5 K (curve 1), 30 K (2), 300 K (3) and 900 K (4). The curve 5 is the correction of (2) using (44). Casimir's result is a dashed line 6

$$\left[\sum_n \tilde{\omega}_n(\kappa)\right]_{f_{e,h}(\tilde{\omega}_n)=0} = \frac{1}{2\pi i}\oint_C \frac{f'_{e,h}(\omega)}{f_{e,h}(\omega)}\omega d\omega = \frac{-1}{2\pi}\oint_C \left[\ln(f_e(\omega))+\ln(f_h(\omega))\right]d(i\omega) =$$

$$= \frac{c}{2\pi}\int_{-\infty}^{\infty}\left[\ln(f_e(\kappa,k,d))+\ln(f_h(\kappa,k,d))\right]dk = \frac{c}{\pi}\int_0^{\infty}\left[\ln(f_e(\kappa,k,d))+\ln(f_h(\kappa,k,d))\right]dk .$$

At the same time, this sum is real, since the poles are complex conjugates, and their imaginary parts cancel each other out. Multiplying by $\hbar$, we obtain the free energy, and differentiating with respect to *d* yields the force (11). The integral over the right infinite semicircle in the complex frequency plane vanishes. For any finite *d*, the squares of the RC tend to zero for large *k*, so the integral over the infinite semicircle vanishes for any finite *d*. However, for perfectly conducting screens, $r_{e,h}^2 = 1$, and the integral (11) diverges when *d*=0. In this case, it diverges logarithmically. For RCs corresponding to real plates described by the plasma model or the Drude model, logarithmic divergence also occurs. Assuming $r_{e,h}^2 = 1$, we obtain the Casimir result for *d*>0 [21,25]. This result should be considered for metals at large *d*. As for formula (12), for ideal conductivity $r_{e,h}^2 = 1$, and it follows

$$P(d,T) = -\frac{2k_B T}{\pi\delta_T^3}\left(\frac{\varsigma(3)\delta_T^3}{8d^3} + \delta_T^3\sum_{n=1}^{\infty}\sum_{m=1}^{\infty}\exp(-x_{nm})\left[\frac{k_n^3}{x_{nm}} + \frac{2k_n^3}{x_{nm}^2} + \frac{2k_n^3}{x_{nm}^3}\right]\right).$$

In this formula $x_{nm} = 2mk_n d$. The result of the calculation using this formula is shown in Fig. 3. Curve 5 corresponds to taking into account a contour integral with a cotangent [26] in addition to (12). In the far zone, it is sufficient to take into account the first term and neglect the next one, i.e. we have a well-known result $P(d,T) = -k_B T\varsigma(3)/(4\pi d^3)$ [3, 31] with dependence $P(d,T) \sim 1/d^3$. In this case, the contour integral is small, and formula (12) gives the correct result [31]. Formula (12) should follow from (11) if, instead of the zero-point energy, we take the average energy of the quantum oscillator $\hbar\omega\coth(\hbar\omega/(2k_B T))/2$ and transform the contour integral as indicated above. Poles appear on the imaginary axis for the cotangent, and the contour integral of the cotangent is taken in the sense of the principal value. It is small at high temperature, but as the temperature approaches zero, it makes a significant contribution — i.e., (12) does not follow from (11) [26] and the result of work [31]. As for formulas (11) and (13), they were originally presented in [25] as an integral over frequency and a variable $\kappa = \sqrt{k_x^2 + k_y^2}$ in the form (see also formula (21.6) from [2])

$$P(d,T) = -\frac{\hbar}{2\pi^2}\int_0^{\infty} d\omega \coth(x)\,\mathrm{Re}\int_0^{\infty}\kappa k_z \sum_{\alpha=e,h}\frac{r_\alpha^2\exp(-2ik_z d)}{1 - r_\alpha^2\exp(-2ik_z d)}d\kappa,$$

where $k_z = \sqrt{k_0^2 - \kappa^2} = -i\tilde{k}_z$, $\tilde{k}_z = \sqrt{\kappa^2 - k_0^2}$ (for $\tilde{\varepsilon} = 1$). Next, a substitution $p = k_z$ is made (formula (1.10) from [25]), and it is claimed that the integral is transformed into the form (11) and (13) without the contours crossing the poles. However, in this integral, the real part is taken, so it is not an analytic function. Reference [25] states that it diverges with frequency. Indeed, this integral is not suitable for calculating the force. Let $r_{e,h}^2 = 1$. Taking the real part, we obtain the integral

$$I_{\alpha 1}(k_0, d) = \int_0^{k_0} \kappa k_z \frac{r_\alpha^2 \exp(-2ik_z d)}{1 - r_\alpha^2 \exp(-2ik_z d)} d\kappa = -k_0^2 / 6 ,$$

and the original integral does not depend on *d* and diverges with frequency. The integral in the region $k_0 < \kappa < \infty$ is imaginary and must be discarded. Physically, this is clear from the fact that this integral corresponds to surface plasmon polaritons (SPPs), which are not supported by ideal screens. The influence of SPPs on the force is considered in [32]. The divergent integral, on the other hand, is associated with radiated modes, which were taken into account in [1].

## 1. Interband conductivity of graphene

In the approximation of the band structure $\varepsilon = \pm\hbar q v_F$, we have interband conductivity (4). A more accurate band structure of graphene is described as $\varepsilon = \pm\gamma_0 w(q_x, q_y)$, and

$$w(\mathbf{q}) = \sqrt{1 + 4\cos(\sqrt{3} q_x a/2)\cos(q_y a/2) + 4\cos^2(q_y a/2)} .$$

In this formula $a = 0.246$ nm is the lattice constant, $\gamma_0 = \pm\hbar q v_F$, and $\hbar\mathbf{q}$ is the quasi-momentum $q = \sqrt{q_x^2 + q_y^2}$. We take the coupling constant $\gamma_0 = 3.0$ eV and the Fermi velocity $v_F = \sqrt{3}\gamma_0 a/(2\hbar) = 0.985 \cdot 10^6$ m/s. For a small photon momentum $p = k_0\hbar = \hbar\omega/c << 2\varepsilon/v_F$, denoting its transverse part $p_\tau\hbar = k_\tau\hbar = q\hbar$, we have

$$f_{FD}(\varepsilon - q\hbar v_F/2) = f_{FD}(\varepsilon) - \partial_\varepsilon f_{FD}(\varepsilon) q\hbar v_F/2 ,$$

$$\partial_\varepsilon f_{FD}(\varepsilon) = -f_{FD}^2(\varepsilon)/(k_B T) = -\partial_{-\varepsilon} f_{FD}(\varepsilon) = \partial_{-\varepsilon} f_{FD}(-\varepsilon),$$

$$f_{FD}(-\varepsilon + q\hbar v_F/2) = f_{FD}(-\varepsilon) + \partial_\varepsilon f_{FD}(\varepsilon) q\hbar v_F/2 .$$

We convert the fractions using an approximate formula

$$\frac{1}{(\omega - i\omega_c \mp q v_F/2)^2 - 4\varepsilon^2/\hbar^2} \approx$$
$$\approx \frac{\hbar^2}{\hbar^2(\omega - i\omega_c)^2 - 4\varepsilon^2} \pm \frac{\hbar^4(\omega - i\omega_c) q v_F}{\left[\hbar^2(\omega - i\omega_c)^2 - 4\varepsilon^2\right]^2} .$$

This is true when $\left|4\varepsilon^2 - \hbar^2\omega^2\right| >> \omega\hbar^2 q v_F = \omega^2\hbar^2 v_F / c$. At low energies, this is true because $v_F / c = 1/300$. This is also true when $\varepsilon >> \hbar\omega$. The violation occurs in the region $\varepsilon \approx \hbar\omega/2$. To first order, denoting the round bracket in (4) as *F*, we obtain

$$F=\hbar^2\frac{f_{FD}(-\varepsilon)-f_{FD}(\varepsilon)}{\hbar^2(\omega-i\omega_c)^2-4\varepsilon^2}+\hbar^4\frac{[f_{FD}(-\varepsilon)+f_{FD}(\varepsilon)](\omega-i\omega_c)qv_F}{\left[\hbar^2(\omega-i\omega_c)^2-4\varepsilon^2\right]^2}+ +\hbar^3\frac{\partial_\varepsilon f_{FD}(\varepsilon)qv_F}{\hbar^2(\omega-i\omega_c)^2-4\varepsilon^2}. \quad (15)$$

It is convenient to denote indefinite integrals

$$\tilde{F}(\varepsilon,\varepsilon_p)=\int\frac{d\varepsilon}{\varepsilon^2-\left(\hbar(\omega-i\omega_c)-\varepsilon_p\right)^2/4}= \frac{1}{\hbar(\omega-i\omega_c)-\varepsilon_p}\ln\left(\frac{\varepsilon-\left[\hbar(\omega-i\omega_c)-\varepsilon_p\right]/2}{\varepsilon+\left[\hbar(\omega-i\omega_c)-\varepsilon_p\right]/2}\right),$$

$$\tilde{\tilde{F}}(\varepsilon)=\int\frac{d\varepsilon}{\left[\varepsilon^2-\hbar^2(\omega-i\omega_c)^2/4\right]^2}= =\frac{-1}{\hbar^2(\omega-i\omega_c)^2}\left(\frac{1}{\varepsilon-\hbar(\omega-i\omega_c)/2}+\frac{1}{\varepsilon+\hbar(\omega-i\omega_c)/2}\right)- -\frac{2}{\hbar^3(\omega-i\omega_c)^3}\ln\left(\frac{2\varepsilon-\hbar(\omega-i\omega_c)}{2\varepsilon+\hbar(\omega-i\omega_c)}\right).$$

The second integral is convenient to use at low temperatures. For $k_BT<<\mu_c$ the first term in (15), it corresponds to the contribution (7). In the second term $f_{FD}(-\varepsilon)=0$, $f_{FD}(\varepsilon)=0$ if $\varepsilon>\mu_c$, and $f_{FD}(\varepsilon)=1$ if $\varepsilon<\mu_c$. In the third term $\partial_\varepsilon f_{FD}(\varepsilon)\approx\delta(\varepsilon-\mu_c)$. Its contribution is

$$\sigma_3^{\text{inter}}(\omega,\omega_c,q,\mu_c,T)=\frac{1}{\pi}\frac{ie^2\hbar(\omega-i\omega_c)qv_F}{\hbar^2(\omega-i\omega_c)^2-4\mu_c^2}. \quad (16)$$

It is capacitive at high frequencies, and at low frequencies it is inductive. The contribution of the second term is equal to

$$\sigma_3^{\text{inter}}(\omega,\omega_c,q,\mu_c,T)=\frac{-ie^2qv_F}{16\pi}\times \times\left(\frac{1}{\hbar(\omega-i\omega_c)/2-\mu_c}-\frac{1}{\hbar(\omega-i\omega_c)/2+\mu_c}-\frac{2}{\hbar(\omega-i\omega_c)}\left(\ln\left(\frac{2\mu_c-\hbar(\omega-i\omega_c)}{2\mu_c+\hbar(\omega-i\omega_c)}\right)-i\pi\right)\right). \quad (17)$$

At a high frequency, the contribution (17) takes the form

$$\sigma_3^{\text{inter}}(\omega,\omega_c,q,\mu_c,T)=\frac{-ie^2\mu_cqv_F}{\pi\hbar^2(\omega-i\omega_c)^2}. \quad (18)$$

At $\mu_c=\hbar\omega$ it is equal

$$\sigma_3^{\text{inter}}(\omega,\omega_c,q,\mu_c,T)=\frac{-ie^2qv_F}{8\pi}\left(\frac{i}{\hbar\omega_c}-\frac{1}{4\mu_c-i\omega_c\hbar}-\frac{1}{\mu_c-i\omega_c\hbar}\left(\ln\left(\frac{\mu_c+i\omega_c\hbar}{3\mu_c-i\omega_c\hbar}\right)-i\pi\right)\right),$$

i.e., at a low CF it is almost real and large: $\sigma_3^{\text{inter}} = e^2 k_0 v_F /(8\pi\hbar\omega_c)$, and at $2\mu_c << \hbar\omega_c$ it is practically zero. At a low frequency, the first-order term (17) contributes $\sigma_3^{\text{inter}} = e^2 q v_F /[8\hbar(\omega - i\omega_c)]$.

For formula (7) to be valid, the condition $\hbar\omega < \mu_c$ must be satisfied. If $\hbar\omega >> \mu_c$, formula (7) yields $\sigma^{\text{inter}}(\omega, \omega_c, \mu_c, T) \approx \sigma_0$, which is fundamentally incorrect: the conductivity must decrease with frequency. This means that the approximation used in (7) is not valid. At low frequency, the contribution of (7) to the conductivity is small:

$$\sigma^{\text{inter}}(\omega, \omega_c, \mu_c, T) \approx \frac{i\sigma_0}{\pi} \frac{(\omega - i\omega_c)\hbar}{|\mu_c|},$$

and this term is of the capacitive type, whereas the intraband conductivity is inductive. When $\mu_c >> \omega\hbar$, the contribution of this term is also small. We obtain the following result: the contribution from interband conductivity at low frequencies is always small and less than the contribution $\sigma^{\text{intra}}$.

The formulas presented above were derived under the condition that $k_B T$ is significantly less than $\mu_c$ and the photon energy. Therefore, the conductivity values presented above do not depend on temperature. The BZ of graphene is of the order $3\gamma_0 = 9$ eV for each of the subbands $\pi^{\pm}$. The photon energy must be finite and must not exceed the band gap. Otherwise, photoionization of graphene occurs, the bonds break, and graphene should be described using the plasma model as a two-dimensional electron gas [16]. Even at $\omega\hbar = \gamma_0 = 3$ eV (i.e., in the near-UV range), the breaking of the π-bond is possible. Since the integral is quite complex, we will approximate it using the mean value theorem. We will replace the infinite upper limit with $\gamma_0$. We will divide the integration interval into three parts: $0 < \varepsilon < \varepsilon_p$, $\varepsilon_p < \varepsilon < \varepsilon_p + \mu_c$, $\varepsilon_p + \mu_c < \varepsilon < \gamma_0$, where $\varepsilon_p = \hbar q v_F / 2$. In the first interval, the FDF varies from $[\exp((-\varepsilon_p - \mu_c)/(k_B T)) + 1]^{-1}$ to $[\exp(-\mu_c/(k_B T)) + 1]^{-1}$ and has a mean value.

$$\bar{f}_1^+ = \frac{[\exp((-\varepsilon_p - \mu_c)/(k_B T)) + 1]^{-1} + [\exp(-\mu_c/(k_B T)) + 1]^{-1}}{2}.$$

Similarly, the function $f_{FD}(-\varepsilon)$ has an average value

$$\bar{f}_1^- = \frac{[\exp((\varepsilon_p - \mu_c)/(k_B T)) + 1]^{-1} + [\exp((2\varepsilon_p - \mu_c)/(k_B T)) + 1]^{-1}}{2}.$$

In other intervals, the functions have the following average values:

$$\bar{f}_2^+ = \frac{1/2 + [\exp(-\mu_c/(k_B T)) + 1]^{-1}}{2},$$

$$\bar{f}_2^- = \frac{\left[\exp\left(-2\mu_c/(k_BT)\right)+1\right]^{-1} + \left[\exp\left(\left(2\varepsilon_p-\mu_c\right)/(k_BT)\right)+1\right]^{-1}}{2},$$

$$\bar{f}_3^+ = \frac{1/2 + \left[\exp\left(\left(\gamma_0-\varepsilon_p-\mu_c\right)/(k_BT)\right)+1\right]^{-1}}{2},$$

$$\bar{f}_3^- = \frac{\left[\exp\left(\left(-\gamma_0+\varepsilon_p-\mu_c\right)/(k_BT)\right)+1\right]^{-1} + \left[\exp\left(\left(2\varepsilon_p-\mu_c\right)/(k_BT)\right)+1\right]^{-1}}{2}.$$

Now integral (4) takes the form

$$\begin{aligned}\sigma^{\text{inter}}(\omega,\omega_c,q,\mu_c,T) &= \frac{-ie^2(\omega-i\omega_c)}{4\pi}\times \\ \times\Big\{&\bar{f}_1^+\left[\tilde{F}(\varepsilon_p,\varepsilon_p)-\tilde{F}(0,\varepsilon_p)\right]-\bar{f}_1^-\left[\tilde{F}(\varepsilon_p,-\varepsilon_p)-\tilde{F}(0,-\varepsilon_p)\right]+ \\ +&\bar{f}_2^+\left[\tilde{F}(\varepsilon_p+\mu_c,\varepsilon_p)-\tilde{F}(\varepsilon_p,\varepsilon_p)\right]-\bar{f}_2^-\left[\tilde{F}(\varepsilon_p+\mu_c,-\varepsilon_p)-\tilde{F}(\varepsilon_p,-\varepsilon_p)\right]+ \\ +&\bar{f}_3^+\left[\tilde{F}(\gamma_0,\varepsilon_p)-\tilde{F}(\varepsilon_p+\mu_c,\varepsilon_p)\right]-\bar{f}_3^-\left[\tilde{F}(\gamma_0,-\varepsilon_p)-\tilde{F}(\varepsilon_p+\mu_c,-\varepsilon_p)\right]\Big\}\end{aligned}. \qquad (19)$$

In formula (19), the quantities $\bar{f}_n^\pm$ depend on temperature, and overall it approximates integral (4) quite accurately for different frequencies and chemical potentials, which is confirmed by numerical calculations. Using formula (19) to calculate integral (11) requires a transition to an imaginary frequency, i.e., to the substitution $i\omega = ik_0c \to kc$, with $k_c = \omega_c/\text{c}$. Introducing the dependence on q takes into account the SD. However, formula (4) is isotropic. In reality, the SD is obtained by integrating over the entire Brillouin zone. It is not feasible to calculate such an integral analytically, and numerically determining it when integrating in (11) requires enormous computational resources. Conductivity without taking into account SD in (19) is obtained when $\varepsilon_p = 0$. If $\omega = 0$, then and $q = 0$, transitions between zones $\pi^\pm$ are impossible, and $\sigma^{\text{inter}} = 0$. Expanding (19) in terms of small parameters, we obtain $\sigma^{\text{inter}}(\omega,\omega_c,q,\mu_c,T) = \sigma_k^{\text{inter}}(\mu_c,T)k + \sigma_\kappa^{\text{inter}}(\mu_c,T)q$.

## 2. The Casimir force for two sheets of graphene at the finite temperature

The Casimir pressure at $T$=0 is calculated using formula (11), in which the RCs are defined for two sheets of graphene on a substrate from the side of the latter. At low temperatures, this formula can be used, in which the conductivity depends on the temperature. In the case of an ideal dielectric substrate, we replace $\varepsilon(k_n,p)\to\tilde{\varepsilon}$ and and the DP can be taken outside the integral sign. We obtain (14) in the form

$$P_{e,h}(d) = -\frac{\hbar c}{32\pi^2 d^4}\int_1^\infty \frac{dp}{p^2}\int_0^\infty \frac{\varepsilon^{-1/2}(p,y)y^3}{r_{e,h}^{-2}(p,y)\exp(y)-1}dy.$$

Since formula (11) was obtained by a Wick rotation, the corresponding substitution should be made in formula (5) for the intraband conductivity, after which it takes the form

$$\sigma^{\text{intra}}(k,T)=\frac{\sigma(0,\mu_c,T)}{1+k/k_c},$$

where $\sigma(0,T)=4\sigma_0 k_B T\varphi(\mu_c,T)/(\pi\hbar k_c c)$, $k_c=\omega_c c$, $k=i\omega/c$. With this substitution, the intraband conductivity becomes a function of k, $\mu_c$ and temperature. The total conductivity now takes the form

$$\sigma(k,\kappa,\mu_c T)=\frac{\sigma(0,\mu_c,T)}{1+k/k_c}+\sigma_\kappa^{\text{inter}}(\mu_c,k,T)v_F\kappa\,, \tag{20}$$

The value $\sigma_\kappa^{\text{inter}}(\mu_c,k,T)$ determines the SD and can be found from (19) as $v_F^{-1}\partial_q\sigma^{\text{inter}}(\omega,\omega_c,q,\mu_c,T)_{q=0}$ when $\omega\to -ikc$. However, when $k_B T<<\mu_c$, this function can be obtained from (4) approximately in the form

$$\sigma_\kappa^{\text{inter}}(\mu_c,k,T)=\frac{\sigma_0(k+k_c)\hbar^2/\pi}{\mu_c^2+((k+k_c)v_F\hbar/2)^2}\,. \tag{21}$$

Since the main contribution comes from low frequencies, the second term in the denominator of (21) can be neglected, and then the interband contribution contains a constant term and a term linear in *k*. The intraband conductivity can also be represented in the form $\sigma^{\text{intra}}(k,T)\approx\sigma(0,\mu_c,T)(1-k/k_c)$. Accordingly, the total conductivity has the form $\sigma(k,T)\approx\sigma_0(k,T)+\sigma_\kappa(k,T)v_F\kappa$. With the substitution $\kappa^2=k^2(p^2-1)$, the conductivity becomes a function of *p*, which takes into account the SD. In deriving (21), we neglected the integral.

$$\frac{e^2 v_F c^2(k+k_c)^2}{\pi}\frac{\hbar^2}{16}\int_0^\infty\frac{(f_{FD}(-\varepsilon)+f_{FD}(\varepsilon))d\varepsilon}{(\varepsilon^2+c^2(k+k_c)^2\hbar^2/4)^2}\,,$$

which can be easily calculated for $T\to 0$, assuming $f_{FD}(-\varepsilon)=1$, $f_{FD}(\varepsilon)=0$ for $\varepsilon>\mu_c$, $f_{FD}(\varepsilon)=1$ for $\varepsilon<\mu_c$, and using formula (120.2) from [33]. We do not present the result, as it is proportional to the second power of $k+k_c$. At low temperatures, the intraband conductivity ceases to depend on it: $\sigma(0,T)=4\sigma_0\mu_c/(\pi\hbar k_c c)$. The possible dependence is related only to CF. We represent the total conductivity in terms of the dimensionless normalized conductivity $\xi$ in the form

$$\sigma(k,T)=\sigma^{\text{intra}}(k,T)+\sigma^{\text{inter}}(k,T)=\xi(k,T)/\sqrt{\varepsilon_0/\mu_0}\,, \tag{22}$$

where $\sqrt{\mu_0/\varepsilon_0}=120\pi$ Ohm. Moving on to summation over Matsubara frequencies in (12), we make the substitution $k\to k_n=i\omega_n/c$, i.e., we use $\xi(k_n,T)$ and formula (12). Accounting for SD in the model under consideration leads to isotropic conductivity. The actual conductivity of graphene is anisotropic [4] (see, for example, (8)–(10)), but the anisotropy is small at low frequencies. It manifests itself at wavelengths comparable to the period of the graphene crystal lattice (less than 1 nm). However, for such short wavelengths, the contribution to the dispersive force is practically absent, since $\xi$ decreases strongly with frequency, and the PD manifests itself mainly as a tensor nature of conductivity. This complicates the recording of the RO, as a dependence on the polarization of the waves emerges.

In conducting graphene sheets, PP modes are possible. Slow, strongly damped PP modes mainly contribute to the force at short distances, while fast, radiated PP modes do so at long distances [32]. At short distances, a more accurate model is based on van der Waals forces. In Fig. 1, it is obtained using density functional theory with approximations based on the Lennard-Jones method and the Grimme method [34]. The temperature dependence of the force is important in the far field. Therefore, in what follows, we consider only scalar conductivity. We introduce normalized conductivities for E-waves $y_e(\varepsilon)=\tilde{\varepsilon}k_0/\sqrt{k_0^2\varepsilon-k_x^2-k_y^2}$ and H-waves $y_h(\varepsilon)=\sqrt{k_0^2\varepsilon-k_x^2-k_y^2}/k_0$ for the substrate with DP $\varepsilon$ and for vacuum $Y_{in}=y_{e,h}(1)$. The input conductivity from the vacuum side to the graphene layer is $Y_{in}=y_{e,h}(1)$. After the graphene layer, it is equal to $Y_{in}=y_{e,h}(1)+\xi$. The RO from the side of the substrate with DP is defined as

$$r_{e,h}(k)=\frac{y_{e,h}(\tilde{\varepsilon})-Y_{in}}{y_{e,h}(\tilde{\varepsilon})-Y_{in}}=\frac{y_{e,h}(\tilde{\varepsilon})-y_{e,h}(1)-\xi(k)}{y_{e,h}(\tilde{\varepsilon})+y_{e,h}(1)+\xi(k)}. \quad (23)$$

In the case of a substrate with dispersion, it should be replaced $\tilde{\varepsilon}\to\varepsilon(\omega)$. Note that for the structure shown in Fig. 2, insert *b* with a vacuum gap, the formula for RC takes the form

$$r_{e,h}(k)=\frac{y_{e,h}(1)-y_{e,h}(\varepsilon(k))-\xi(k)}{y_{e,h}(1)+y_{e,h}(\varepsilon(k))+\xi(k)}. \quad (24)$$

In (23) and (24), we switched to the imaginary frequency, i.e. $y_e(\tilde{\varepsilon})=\tilde{\varepsilon}k/\sqrt{k_x^2+k_y^2+k^2\tilde{\varepsilon}}$, $y_h(\tilde{\varepsilon})=\sqrt{k_x^2+k_y^2+k^2\tilde{\varepsilon}}/k$, and used the substitution applied in (12) $\kappa^2=k_x^2+k_y^2=(p^2-1)\tilde{\varepsilon}k^2$, i.e. $y_e(\tilde{\varepsilon},p)=\sqrt{\tilde{\varepsilon}}/p$, $y_h(\tilde{\varepsilon},p)=p\sqrt{\tilde{\varepsilon}}$. For the DP substrate, we have

$$r_{ne}(k_n,p)=\frac{\sqrt{\tilde{\varepsilon}}-1-p\xi(k_n,T)}{\sqrt{\tilde{\varepsilon}}+1+p\xi(k_n,T)}, \quad (25)$$

$$r_{nh}(k_n,p)=\frac{p(\sqrt{\tilde{\varepsilon}}-1)-\xi(k_n,T)}{p(\sqrt{\tilde{\varepsilon}}+1)+\xi(k_n,T)}. \quad (26)$$

In the approximation of intraband conductivity, we obtain

$$r_e(k_n,p)=\frac{(\sqrt{\tilde{\varepsilon}}-1)(k_c+k_n)-pk_c\xi(0,T)}{(\sqrt{\tilde{\varepsilon}}+1)(k_c+k_n)+pk_c\xi(0,T)}, \quad (27)$$

$$r_h(k_n,p)=\frac{p\left(\sqrt{\tilde{\varepsilon}}-1\right)(k_c+k_n)-k_c\xi(0,T)}{p\left(\sqrt{\tilde{\varepsilon}}+1\right)(k_c+k_n)+k_c\xi(0,T)}. \quad (28)$$

Here $\xi(0,T)=\sigma(0,T)\sqrt{\mu_0/\varepsilon_0}$. Taking into account the SD requires supplementing term (21). We consider relations (11)–(14) in the form $P(d,T)=P_e(d,T)+P_h(d,T)$, where for (12)

$$P_{e,h}(d,T)=-\frac{k_B T\tilde{\varepsilon}^{3/2}}{\pi}\sum_{n=0}^{\infty}\frac{k_n^3}{1+\delta_{n0}}\int_1^{\infty}\sum_{m=1}^{\infty}r_{e,h}^{2m}(p)\exp\left(-2pmk_n\tilde{\varepsilon}^{1/2}d\right)p^2dp. \quad (29)$$

The integrals $I_{(e,h)nm}(d)$ in (29) are calculated using integration by parts:

$$I_{(e,h)nm}(d)=\int_1^{\infty}r_{e,h}^{2m}(p)\exp\left(-2pmk_n\tilde{\varepsilon}^{1/2}d\right)p^2dp=$$

$$=\frac{r_{e,h}^{2m}(1)\exp\left(-2mk_n\tilde{\varepsilon}^{1/2}d\right)}{2mk_n\tilde{\varepsilon}^{1/2}d}+\frac{\partial_p\left(p^2r_{e,h}^{2m}(p)\right)_{p=1}\exp\left(-2mk_n\tilde{\varepsilon}^{1/2}d\right)}{\left(2mk_n\tilde{\varepsilon}^{1/2}d\right)^2}+\ldots+ \ .$$

$$\frac{\partial_p^l\left(p^2r_{e,h}^{2m}(p)\right)_{p=1}\exp\left(-2mk_n\tilde{\varepsilon}^{1/2}d\right)}{\left(2mk_n\tilde{\varepsilon}^{1/2}d\right)^{l+1}}+\ldots$$

The final formula takes the form

$$P_{e,h}(d,T)=-\frac{k_B T\tilde{\varepsilon}^{3/2}}{\pi}\sum_{n=1}^{\infty}\sum_{m=1}^{\infty}I_{(e,h)nm}(d). \quad (30)$$

If we expand $r_{e,h}^{2m}(k_n,p)$ in series in terms of $p$ at zero, we could calculate the integrals in (28) analytically. However, unfortunately, such expansions do not converge for $p>1$, although formally the exponential ensures the convergence of the double series in n and m. A good approximation to (30) in the far zone is already given by the formula.

$$P_{e,h}(d,T)\approx-\frac{k_B T\tilde{\varepsilon}^{3/2}}{\pi}I_{(e,h)11}(d). \quad (31)$$

To calculate the initial coefficients, we have the following relationships:

$$\partial_p^0\left(p^2r_{e,h}^{2m}(k_n,p)\right)_{p=1}=r_{e,h}^{2m}(1),$$

$$\partial_p^1\left(p^2r_{e,h}^{2m}(k_n,p)\right)_{p=1}=2r_{e,h}^{2m}(1)+\partial_p r_{e,h}^{2m}(1),$$

$$\partial_p^2\left(p^2r_{e,h}^{2m}(k_n,p)\right)_{p=1}=3\partial_p r_{e,h}^{2m}(1)+2\partial_p^2 r_{e,h}^{2m}(1),$$

$$\partial_p^3\left(p^2r_{e,h}^{2m}(k_n,p)\right)_{p=1}=6\partial_p^2 r_{e,h}^{2m}(1)+2\partial_p^3 r_{e,h}^{2m}(1),$$

$$r_{ne}^{2m}(k_n,1)=\left(\frac{\sqrt{\tilde{\varepsilon}}-1-\xi(k_n,T)}{\sqrt{\tilde{\varepsilon}}+1+\xi(k_n,T)}\right)^{2m},$$

$$\partial_p r_{ne}^{2m}(k_n,1)=-4mr_{ne}^{2m-1}(k_n,1)\xi(k_n,T)\frac{1+\xi(k_n,T)}{\left[\sqrt{\tilde{\varepsilon}}+1+\xi(k_n,T)\right]^2},$$

$$r_{nh}^{2m}(k_n,1)=\left(\frac{\sqrt{\tilde{\varepsilon}}-1-\xi(k_n,T)}{\sqrt{\tilde{\varepsilon}}+1+\xi(k_n,T)}\right)^{2m},$$

$$\partial_p r_{nh}^{2m}(k_n,1)=2mr_{nh}^{2m-1}(k_n,1)\frac{\sqrt{\tilde{\varepsilon}}\left(\sqrt{\tilde{\varepsilon}}+1\right)}{\left[p\left(\sqrt{\tilde{\varepsilon}}+1\right)+\xi(k_n,T)\right]^2}.$$

For the vacuum gap $\tilde{\varepsilon}=1$, the relations (24) and (25) simplify. Denoting $x_n=\xi(k_n,T)/2$, $y_n=1/x_n=2/\xi(k_n,T)$,, we then have the formal expansions:

$$r_{ne}^{2m}(k_n,p)=\frac{x_n^{2m}p^{2m}}{(1+x_np)^{2m}}=x_n^{2m}\left[p^{2m}-2mx_np^{2m+1}+m(2m-1)x_n^2p^{2m+2}-...\right], \tag{32}$$

$$r_{nh}^{2}(k_n,p)=\frac{x_n^{2m}}{p^{2m}(1+x_n/p)^{2m}}=\left[\frac{x_n^{2m}}{p^{2m}}-2m\left(\frac{x_n}{p}\right)^{2m+1}+m(2m-1)\left(\frac{x_n}{p}\right)^{2m+2}-...\right]. \tag{33}$$

These expansions work well when the low-frequency conductivity is small, i.e. $x_0=\xi(0,T)/2<1$, since the values $x_n$ decrease as *n* increases. In the general case, derivatives can be calculated by differentiating the left-hand sides of (32) and (33). In the far zone $d/\delta_T>1$, where it makes sense to consider the temperature correction, the values $a_{nm}=2mk_n\tilde{\varepsilon}^{1/2}d=4\pi nm\tilde{\varepsilon}^{1/2}d/\delta_T$ under the exponent are large for large indexes. Even for indexes of the order of one, the exponent is small, and the terms with the next exponent can already be neglected. Thus, formula (31) provides a good approximation to the solution of the problem. According to (26) and (27), the squares of the RCs are less than one. This means that the conductivity of graphene is finite and the value of the fraction in (12) is finite. If $k_c=0$, the conductivity of graphene at zero frequency becomes infinite (the conductivity model $\sigma(0,T)$ is inversely proportional to $k_c=0$). According to (20) $r_{e,h}(0)=-1$. Then an indeterminacy of 0/0 arises in (12). To reveal it, we represent $\exp\left(2pk_n\tilde{\varepsilon}^{1/2}d\right)\approx 1+2pk_n\tilde{\varepsilon}^{1/2}d$, $k_n\to 0$, and we see that the term with *n*=0 drops out. In the near zone, *d* is small compared to $\delta_T$, and in formula (30) many terms should be taken into account, since the exponentials become close to one. This is the disadvantage of this formula for analytical calculations, since it requires not only many terms of the double series but also many calculations of the derivatives $\partial_p^l\left(p^2r_{e,h}^{2m}(p)\right)_{p=1}$. However, in the near zone, the influence of temperature is weak, and at temperatures close to zero, it is convenient to use formula (11) or (13).

## 3. The Casimir force at low temperature

At low temperatures, we use formulas (11) or (13), in which the low-frequency conductivity is taken with a temperature dependence. For an ideal dielectric substrate, the formulas take the form

$$P_{e,h}(d) = -\frac{\hbar c\tilde{\varepsilon}^{3/2}}{2\pi^2}\sum_{m=1}^{\infty}\int_1^{\infty} dp p^2 \int_0^{\infty} k^3 r_{e,h}^{2m}(p,k)\exp\left(-2mpk\sqrt{\tilde{\varepsilon}}d\right)dk,$$

where, for low frequencies $k < k_c$, and one can take expansions $r_e^{2m}(k,p) = r_e^{2m}(0,p) + \Delta_e(p)k/k_c$, $r_h^{2m}(k,p) = r_h^{2m}(0,p) + \Delta_h(p)k/k_c$ in which

$$r_e^{2m}(0,p) = \frac{\left[\sqrt{\tilde{\varepsilon}} - 1 - p\xi(0,T)\right]^{2m}}{\left[\sqrt{\tilde{\varepsilon}} + 1 + p\xi(0,T)\right]^{2m}},$$

$$\Delta_{em}(p) = \frac{4mp\sqrt{\tilde{\varepsilon}}\xi(0,T)r_e^{2m}(0,p)}{\left(\sqrt{\tilde{\varepsilon}} + 1 + p\xi(0,T)\right)\left(\sqrt{\tilde{\varepsilon}} - 1 - p\xi(0,T)\right)},$$

$$r_h^{2m}(0,p) = \frac{\left[p\left(\sqrt{\tilde{\varepsilon}} - 1\right) - \xi(0,T)\right]^{2m}}{\left(p\left(\sqrt{\tilde{\varepsilon}} + 1\right) + \xi(0,T)\right)^{2m}},$$

$$\Delta_{hm}(p) = \frac{4mp\sqrt{\tilde{\varepsilon}}\xi(0,T)r_h^{2m}(0,p)}{\left[p\left(\sqrt{\tilde{\varepsilon}} + 1\right) + \xi(0,T)\right]\left[p\left(\sqrt{\tilde{\varepsilon}} - 1\right) - \xi(0,T)\right]}.$$

Integration with respect to $k$ is calculated quite simply:

$$\int_0^{\infty}\left(k^3 r_{e,h}^{2m}(0,p) + \Delta_{e,h}(p)k^4\right)\exp\left(-2mpk\sqrt{\tilde{\varepsilon}}d\right)dk =$$
$$= \frac{6r_{e,h}^{2m}(0,p)}{\left(2mp\sqrt{\tilde{\varepsilon}}d\right)^4} + \frac{24\Delta_{(e,h)m}(p)}{\left(2mp\sqrt{\tilde{\varepsilon}}d\right)^5 k_c}.$$

With this decomposition, the force in the far field is defined as $P(d,T) = -\hbar c\left(a_0 + a_1/(k_c d)\right)/\left(2\pi^2\sqrt{\tilde{\varepsilon}}d^4\right)$, where the coefficients have the form

$$a_0 = \sum_{m=1}^{\infty}\int_1^{\infty} p^2\left(r_e^{2m}(0,p) + r_h^{2m}(0,p)\right)dp,$$

$$a_1 = \sum_{m=1}^{\infty}\int_1^{\infty} p\left(\Delta_{em}(p) + \Delta_{hm}(p)\right)dp.$$

These integrals are easier to calculate numerically. However, if $\xi(0,T) << \sqrt{\tilde{\varepsilon}} - 1$, which is usually the case, then for $p < p_{\max} = \left(\sqrt{\tilde{\varepsilon}} - 1\right)/\xi(0,T)$ we obtain, as a first approximation

$$r_e^{2m}(0,p) = r_h^{2m}(0,p) = r^{2m} = \frac{\left(\sqrt{\tilde{\varepsilon}} - 1\right)^{2m}}{\left(\sqrt{\tilde{\varepsilon}} + 1\right)^{2m}},$$

$$\Delta_{em}(p)=\frac{4mp\sqrt{\tilde{\varepsilon}}\,\xi(0,T)r^{2m}}{\left(\sqrt{\tilde{\varepsilon}}+1\right)\left(\sqrt{\tilde{\varepsilon}}-1\right)},$$

$$\Delta_{hm}(p)=\frac{4m\sqrt{\tilde{\varepsilon}}\,\xi(0,T)r^{2m}}{p\left(\sqrt{\tilde{\varepsilon}}+1\right)\left(\sqrt{\tilde{\varepsilon}}-1\right)},$$

$$a_0=(2/3)\left(p_{\max}^3-1\right)\sum_{m=1}^{\infty}r^{2m}=\frac{\left(\sqrt{\tilde{\varepsilon}}+1\right)^2}{6\sqrt{\tilde{\varepsilon}}}\left(\frac{\left(\sqrt{\tilde{\varepsilon}}-1\right)^3}{\xi^3(0,T)}-1\right),$$

$$a_1=\frac{4\xi(0,T)\sqrt{\tilde{\varepsilon}}\left(p_{\max}-1\right)\left(p_{\max}^2+2\right)}{3\left(\sqrt{\tilde{\varepsilon}}+1\right)\left(\sqrt{\tilde{\varepsilon}}-1\right)}\sum_{m=1}^{\infty}mr^{2m}=$$
$$=\frac{4\xi(0,T)\sqrt{\tilde{\varepsilon}}\left(p_{\max}-1\right)\left(p_{\max}^2+2\right)}{3\left(\sqrt{\tilde{\varepsilon}}+1\right)\left(\sqrt{\tilde{\varepsilon}}-1\right)\left(1-r^2\right)^2}.$$

If we consider sheets with infinite conductivity, then for $\tilde{\varepsilon}=1$ the Casimir result [21] $P(d)=-\hbar c\pi^2/\left(240d^4\right)$ immediately follows. The method of reducing to the Casimir formula is based on the substitution $\kappa^2=k^2\tilde{\varepsilon}\left(p^2-1\right)$, $K=pk\sqrt{\tilde{\varepsilon}}$, $\kappa d\kappa=k^2\tilde{\varepsilon}pdp$, and for perfectly conducting sheets on a substrate with DP $\tilde{\varepsilon}$, we obtain $P(d)=-\hbar c\pi^2/\left(240\tilde{\varepsilon}^{1/2}d^4\right)$, i.e., the substrate reduces the force. This can be interpreted as an increase in energy in the gap. Since $r_{e,h}^{-2}(p,y)>1$, the force for graphene is less than for perfectly conducting sheets. For low frequencies, the magnitude $k=y/\left(2p\sqrt{\tilde{\varepsilon}}d\right)$ must be small. This means that $y$ is small when $2p\sqrt{\tilde{\varepsilon}}d>1$. Since $p\sqrt{\tilde{\varepsilon}}>1$ and $k$ is small, the integration it will always be in the limit $d\to\infty$. By switching to the parameter $y$ and taking into account the expansion in $y$, we obtain the contribution to the force in the far field. If we take into account that $r_e^{2m}(0,p)\approx r_h^{2m}(0,p)\approx r^{2m}$ and neglect the decrease $r_{e,h}^2$ with frequency and the corrections, we obtain the force reduction factor.

$$\varsigma^{-1}(4)\sum_{m=1}^{\infty}\frac{r^{2m}}{m^4}.$$

However, the actual reduction is much greater. For the model of normalized graphene conductivity (5) or (20), we introduce the wave number $k_\Omega(T)=\sigma(0,T)/k_c$. Now $\xi(p,y)=2p\sqrt{\tilde{\varepsilon}}dk_\Omega/\left(y+k_c2p\sqrt{\tilde{\varepsilon}}d\right)$. The contribution to the force is mainly due to small frequencies, i.e., small $y$ and large $pd$. Using this, we have $\xi(p,y)\approx\xi(0)-y\xi_1$, $\xi(0)=k_\Omega/k_c$, $\xi_1=1/\left(2\sqrt{\tilde{\varepsilon}}k_cd\right)$. We also have $y_e(\tilde{\varepsilon})=\sqrt{\tilde{\varepsilon}}/p$, $y_h(\tilde{\varepsilon})=p\sqrt{\tilde{\varepsilon}}$. Now $r_{e,h}^{2m}(0,p) \approx r_{e,h}^{2m}(p) - \Delta_{(e,h)m}(p) y / \left(2mp\sqrt{\tilde{\varepsilon}}\, d\right)$. Let us write formula (14) in the form of a series expansion. After integrating over y, it takes the form

$$P_{e,h}(d) = -\frac{\hbar c}{32\pi^2 \tilde{\varepsilon}^{1/2} d^4} \sum_{m=1}^{\infty} \int_1^{\infty} \frac{dp}{p^2} \left( \frac{r_{e,h}^{2m}(0,p)\Gamma(4)}{m^4} - \frac{\Delta_{e,h}(p)}{2m^5 p \tilde{\varepsilon}^{1/2} k_c d} \Gamma(5) \right). \tag{34}$$

The main problem now is to calculate the integrals in (34). They are easily calculated numerically. The simplest way to calculate them is when $\xi(0,T) << 1$. In this case, the integral is equal to

$$\frac{r^{2m}}{m^4} \left( \Gamma(4) - \frac{4\xi(0,T)\Gamma(5)}{d\left(\sqrt{\tilde{\varepsilon}}+1\right)\left(\sqrt{\tilde{\varepsilon}}-1\right)} \right).$$

For an approximate estimation of integrals, the mean value theorem can be applied. Thus, the square $r_e^{2m}(k,p)$ changes from $\left(\xi(k)+1-\sqrt{\tilde{\varepsilon}}\right)^{2m} / \left(\xi(k)+1+\sqrt{\tilde{\varepsilon}}\right)^{2m}$ to unity as $p$ varies . We take the average value.

$$\bar{r}_e^{2m}(k) = \frac{\left(\xi(k)+1+\sqrt{\tilde{\varepsilon}}\right)^{2m} + \left(\xi(k)+1-\sqrt{\tilde{\varepsilon}}\right)^{2m}}{2\left(\xi(k)+1+\sqrt{\tilde{\varepsilon}}\right)^{2m}}.$$

Similarly

$$\bar{r}_h^{2m}(k) = \frac{\left(\sqrt{\tilde{\varepsilon}}-1\right)^{2m} / \left(\sqrt{\tilde{\varepsilon}}+1\right)^{2m} + \left(\xi(k)+1-\sqrt{\tilde{\varepsilon}}\right)^{2m} / \left(\xi(k)+1+\sqrt{\tilde{\varepsilon}}\right)^{2m}}{2}.$$

In the first approximation, after integrating with respect to *k*, we have

$$\left(\bar{r}_e^{2m}(0) + \bar{r}_h^{2m}(0)\right) \frac{\Gamma(4)}{\left(2mp\sqrt{\tilde{\varepsilon}}\, d\right)^4}.$$

After integration by *p*, we have

$$\left(\bar{r}_e^{2m}(0) + \bar{r}_h^{2m}(0)\right) \frac{\Gamma(4)}{\left(2m\sqrt{\tilde{\varepsilon}}\, d\right)^4}$$

Thus

$$P_{e,h}(d) \approx -\frac{\hbar c}{32\pi^2 \tilde{\varepsilon}^{1/2} d^4} \sum_{m=1}^{\infty} \frac{\left(\bar{r}_e^{2m} + \bar{r}_h^{2m}\right)}{m^4}. \tag{35}$$

This is the force in the far field. Since the formula was derived for low frequencies, taking into account the zeroth-order expansion in *k*, it is valid for large distances. The expansions of $\bar{r}_{e,h}^{2m}(k)$ make it possible to obtain corrections.

Let us consider the case of no substrate $\tilde{\varepsilon} = 1$. Then the relations simplify. They are particularly simple for low conductivity of graphene $\xi(0,T) << 1$, for which we have:

$$r_e^{2m}(0,p)=\frac{p^{2m}\xi^{2m}(0,T)}{(2+p\xi(0,T))^{2m}}\approx$$
$$\approx\frac{p^{2m}\xi^{2m}(0,T)}{2^{2m}}(1-mp\xi(0,T)),$$

$$\Delta_{em}(p)=-\frac{4mr_e^{2m}(0,p)}{2+p\xi(0,T)}\approx$$
$$\approx 2mr_e^{2m}(0,p)(1-p\xi(0,T)/2),$$

$$r_h^{2m}(0,p)=\frac{\xi^{2m}(0,T)}{(2p+\xi(0,T))^{2m}}\approx$$
$$\approx\frac{\xi^{2m}(0,T)}{(2p)^{2m}}\left(1-\frac{m\xi(0,T)}{p}\right), \tag{36}$$

$$\Delta_{hm}(p)=-\frac{4mpr_h^{2m}(0,p)}{2p+\xi(0,T)}\approx$$
$$\approx -2mr_h^{2m}(0,p)\left(1-\frac{\xi(0,T)}{2p}\right).$$

The relationships for the electrical mode are obtained at $p<2/\xi(0,T)$. The value $2/\xi(0,T)$ should be used as an upper limit for *p*. Large *p* correspond to sliding incidence of waves, in which H-waves are not reflected, which justifies the truncation by *p*. We will neglect the corrections in $r_{e,h}^{2m}$. Then, after calculating the integrals, the contribution from the electrical modes is equal to (for $m>1$)

$$S_{em}(d)=\frac{\xi^{2m}(0,T)}{2^{2m}m^4}\left[\left(\Gamma(4)+\frac{\xi(0,T)}{2\tilde{\varepsilon}^{1/2}k_c d}\Gamma(5)\right)\frac{p_{\max}^{2m-1}-1}{2m-1}-\frac{\Gamma(5)}{\tilde{\varepsilon}^{1/2}k_c d}\frac{p_{\max}^{2m-2}-1}{2m-2}\right],$$

and for magnetic modes, respectively

$$S_{hm}(d)=\frac{\xi^{2m}(0,T)}{2^{2m}m^4}\left(\frac{\Gamma(4)}{2m-1}-\frac{\Gamma(5)}{\tilde{\varepsilon}^{1/2}k_c d}\left(\frac{1}{2m-2}-\frac{\xi(0,T)}{4m-6}\right)\right).$$

When $m=1$, we should replace $\left(p_{\max}^{2m-2}-1\right)/(2m-2)\to\ln(2/\xi(0,T))$ and $1/(2m-2)\to-\ln(2/\xi(0,T))$. We obtain the final result for the far zone

$$P(d)=-\frac{\hbar c}{32\pi^2\tilde{\varepsilon}^{1/2}d^4}\sum_{m=1}^{\infty}\left(S_{em}(d)+S_{hm}(d)\right). \tag{37}$$

Since the terms in the sum are proportional to $2^{-2m}m^{-4}\xi^{2m}(0,T)$, in (37) we can keep only one term, and then

$$P(d)=-\frac{3\hbar c\xi^2(0,T)}{32\pi^2\tilde{\varepsilon}^{1/2}d^4}\left(1+\frac{\xi(0,T)}{\tilde{\varepsilon}^{1/2}k_c d}(1-\xi(0,T))\right). \tag{38}$$

It is interesting to note that the logarithms have been removed from the result. Result (38) is noticeably smaller in $0.23\xi^2(0)/\tilde{\varepsilon}^{1/2}$ times than Casimir's result. In another limiting case $\xi(0,T) >> 1$, the integrals can also be easily calculated, but this hypothetical case is not feasible.

**4. Давление Казимира в ближней зоне**

The near zone is defined by small d, comparable to the wavelengths considered in the spectrum. For wavelengths on the order of 1 nm and less, the substance exhibits discrete properties, and the van der Waals interaction should be considered. For such wavelengths, graphene is transparent and is not described by conductivity. However, some average force, as a combination of the interaction forces between atoms and molecules, does exist. Here, it is convenient to use quantum mechanics methods, such as density functional theory. Figure 1 shows the results obtained based on density functional theory (DFT) for the interaction energy of graphene sheets in a vacuum. According to Figure 1, the force changes sign at a distance of about 0.34 nm; at the same time, differentiating the energy yields a maximum force density of about $10^9$ N/m² at $d$~0.5 nm, followed by a decrease to $10^7$ N/m² at $d$~1 nm. Already in the near UV range, photoionization of graphene is possible, i.e., its transformation into a two-dimensional electron gas. It should have a significantly lower collision frequency, since electrons cannot move to a significant distance from the atom within a short period. The convergence of the integral at high frequencies in formulas of type (29) is ensured by a large exponential in the denominator. It ensures the decay of integrals of the type (11), (12) on the remote semicircle in the complex plane and the validity of the "argument principle" theorem, on the basis of which the aforementioned formulas are derived. When $d$=0, the conditions of the theorem are not satisfied. In this case, the squares of the ROs, which decay at high frequency, do not ensure convergence. It is possible to demonstrate the logarithmic divergence of the integral in this case. Convergence can be ensured if we assume a faster decay of the conductivity with frequency. This can be ensured by the decay of CF with frequency. A characteristic frequency can be defined as the frequency $\Omega$, at which $\xi = -i\xi_0/(\pi\hbar) = -i0.0226$. From a physical perspective, this model is quite justified. Even for the frequency $4\Omega$, it can be assumed that graphene does not have an effect. At $\mu_c = 0.01$ eV, this corresponds to 40 THz. However, from the same perspective, it is also justified to cut off the integral at a certain limiting frequency, at which conductivity no longer has an effect or does not occur. It can be assumed that CF does not depend on the frequency up to a certain critical value, after which it begins to decrease sharply. As a result, the conductivity decreases according to a law that is more pronounced than $\omega^{-1}$. For numerical calculations at zero temperature, a model was used:

$$\omega_c(\omega)=\frac{\omega_{c0}\omega_p^\nu}{\omega_p^\nu+\omega^\nu}.$$

$$\sigma(\omega)=\frac{\sigma(0)\omega_{c0}\omega_p^\nu}{\omega_{c0}\omega_p^\nu+i\omega\left(\omega_p^\nu+\omega^\nu\right)}. \qquad (39)$$

$$\sigma(0)=4\sigma_0\mu_c/(\pi\hbar\omega_{c0}).$$

In it, $\omega_p$ = 4Ω is the characteristic frequency, and $\omega_{c0}$ is the CF at zero frequency. If we take $\nu=3$ or more, this will ensure convergence for all $d$. The results were obtained at $\nu=4$. At the same time, the results are practically independent of ν, since at low frequencies the conductivity is practically independent of this parameter. In this case, we can expand the exponential into a series $\exp\left(-2pk\sqrt{\varepsilon(k)}d\right)=1-2pk\sqrt{\varepsilon(k)}d+...$ and obtain power-law dependencies of the force in the near field. It should be noted right away that the minimum value of $d$ is on the order of 1 nm, after which the consideration becomes meaningless. For any finite $d$, the integral can be calculated by increasing the upper limit. Moreover, the discarded part can be estimated using integration by parts. When calculating the temperature dependencies for CF, it was taken that $\omega_{c0}(T)=\omega_{c00}T/T_0$, where $\omega_{c00}$ is CF at $T_0=300$ K.

We have considered a substrate without dispersion. Now let us consider a metal substrate with DP dispersion $\varepsilon(k)=\varepsilon_L+k_p^2/\left(k^2+kk_{co}\right)$, where $k_{co}$ is the wavenumber for CF in the metal, as well as metal plates with graphene sheets and without graphene. In particular, at low frequencies, we can take the Drude model in the form of $\varepsilon(k)=k_p^2/(kk_{co})$, or even DP $\varepsilon(k)=\varepsilon_L+k_p^2/k^2$, which corresponds to the absence of CF (plasma model). At extremely high frequencies, the Lorentz term $\varepsilon_L$ tends to unity. The models differ only in the ROs, which for configuration $b$, Fig. 2, are determined from the side of the vacuum gap using formula (23), and the results depend only on the expansions of the ROs squares. Since the surface conductivity of the metal at low frequencies is significantly higher than that of graphene, graphene sheets have little effect on the force, which is close to the Casimir result. In the case of graphene on a plasma layer and on a metal substrate (curves 3, 4), the pressure decreases, and in the far field it decreases according to an exponential-power law, i.e., faster than according to the law $1/d^4$ (curve 4). This can be explained by the fact that at low frequencies, the field almost does not penetrate into the metal (including zero-point vacuum oscillations), and, on the other hand, the energy of the oscillations is proportional to the DP. The law for the pressure on two graphene sheets on a metal substrate at large d is obtained from the low-frequency approximation for the DP of the metal $\varepsilon(k)=k_p^2/(kk_{c0})$, which is valid at low frequencies. Then $y=2pk_pd\sqrt{k/k_{co}}$, $r_{e,h}^2 = 1$, $k = y^2 k_{co} / \left(4p^2 k_p^2 d^2\right)$, $dk = k_{co} y dy / \left(2p^2 k_p^2 d^2\right)$, and we obtain, considering CF to be very small and the approximation $r_{e,h}^2 = 1$ to be applicable,

$$P(d) \approx -\frac{\hbar c k_{co}}{32\pi^2 k_p^2 d^5} \int_1^\infty \frac{dp}{p^2} \int_0^\infty \frac{y^4}{\exp(y) - 1} dy = -\frac{\hbar c k_{co} \pi^2}{1440 k_p^2 d^5}. \tag{40}$$

The actual force is somewhat less, since $r_{e,h}^2 < 1$. It can be shown that the influence of graphene in this case is reduced to a correction of the order of $1/d^6$, i.e., (40) determines the pressure on the metal layer. It can be interpreted as a macroscopic result of the van der Waals interaction between metal atoms. It is also possible to refine the coefficient in (40), taking into account that $r_{e,h}^2(0) < 1$. Specifically, the integrals take the form

$$r_{e,h}^2(0) \sum_{m=1}^\infty \int_0^\infty y^4 \exp(-my) dy = \frac{r_{e,h}^2(0)\pi^2}{90},$$

and in (40) a multiplier $\left(r_e^2(0) + r_h^2(0)\right)/2$ appears. In the near zone, when $k_p d = 1$ (i.e. $d = \lambda_p / (2\pi) \sim 20$ nm), the force (40) is approximately in $k_{c0} / \left(3k_p\right) \sim 6 \cdot 10^3$ times smaller than the Casimir force. The DP model $\varepsilon(k) = k_p^2 / k^2$ without collisions cannot be used, since at high frequencies $\varepsilon(\infty) = 0$, and the integral over *k* diverges. A more realistic model $\varepsilon(k) = 1 + k_p^2 / k^2$ should be used. For it, we have $y = 2pk\sqrt{1 + k_p^2 / k^2}\, d$, $\kappa^2 = k^2\left(1 + k_p^2 / k^2\right)\left(p^2 - 1\right)$, $\sqrt{k^2 \varepsilon(k) + \kappa} = k\sqrt{\varepsilon(k)}\, p$, and the dependencies.

$$r_e(k, p) = \frac{\sqrt{\varepsilon} / p - 1/\sqrt{\left(p^2 - 1\right)\varepsilon} - \xi(0)/(k + k_c)}{\sqrt{\varepsilon} / p + 1/\sqrt{\left(p^2 - 1\right)\varepsilon} + \xi(0)/(k + k_c)}, \tag{41}$$

$$r_h(k) = \frac{\sqrt{\varepsilon}\left(p - \sqrt{p^2 - 1}\right) - k\xi(0)/(k + k_c)}{\sqrt{\varepsilon}\left(p + \sqrt{p^2 - 1}\right) + k\xi(0)/(k + k_c)}. \tag{42}$$

At low frequencies, the DP is large, and the squares of these coefficients are close to one. Using their unit values in (13), we obtain

$$P_{e,h}(d) = -\frac{\hbar c}{2\pi^2} \int_1^\infty dp p^2 \sum_{m=1}^\infty \int_0^\infty k^3 \varepsilon^{3/2}(k) \exp\left(-2mpk\sqrt{\varepsilon} k d\right) dk.$$

Integrating by parts three times with respect to *p*, we have

$$P_{e,h}(d) = -\frac{\hbar c}{2\pi^2} \sum_{m=1}^\infty \frac{1}{(2md)^l} \sum_{l=1}^3 \int_{k_p}^\infty \exp(-2mxd) \frac{x^{4-l} dx}{\sqrt{x^2 - k_p^2}}. \tag{43}$$

Result (43) shows that the pressure decreases according to a complex exponential-power law. The integral in (43) can be estimated by dividing the domain into two parts: $k_p < x < 2k_p$, and $2k_p < x < \infty$.. The main contribution to the integral comes from the first domain. In it, we can take the average value of the exponential function $\exp(-2mk_p d)(1+\exp(-2mk_p d))/2 \approx \exp(-2mk_p d)/2$, since $k_p d > 1$. As a result, the integrals can be calculated. For example, for $l$=1, the integral is equal to $k_p^3 \exp(-2mk_p d)[1/\sqrt{3}+\sqrt{3}]/2$. The remainder of the integral is estimated using integration by parts, and it is small. As a result, when the exponent $\exp(-2mk_p d)$ is applied, terms arise $k_p^3/d$, $k_p^2/d^2$, $k_p/d^3$. There is no summation formula for the series in terms of $m$, but its convergence is extremely high, and the main contribution is made by the first term.

The results (40), (43) were obtained when the ROs ate close to one. The conductivity of graphene was excluded from them, since it is limited, while for a metal it is infinite at zero frequency. However, for sufficiently high frequencies, the ROs tend to zero. This leads to an increase in the denominators in formula (11) and to a decrease in the real strength compared to (40). The appearance of terms corresponding to the influence of graphene is related to taking into account the following terms in the expansions of the squares of the ROs.

For configuration $b$, Fig. 2, graphene has a negligible effect on the result. Without taking it into account, in the metal DP approximation $\varepsilon(k) = k_p^2/k^2$, accounting for the third-order RCs expansions yields an expansion of the Casimir force up to $(d/\delta_p)^{-3}$:

$$P(d) = -\frac{\hbar c \pi^2}{240 d^4}\left(1 - \frac{a_1}{d/\delta_p} + \frac{a_2}{(d/\delta_p)^2} + \frac{a_3}{(d/\delta_p)^3} + \ldots\right) \tag{44}$$

with the following coefficients [27]: $a_1 = 16/3$, $a_2$=24, as well as

$$a_3 = 10\varsigma(6)/\varsigma(4) + 1950\varsigma(5) - 1200\varsigma(6)/(7\varsigma(4)).$$

The coefficient $a_1$ matches the results from [28,35]. Using the DP model $\varepsilon = 1 + k_p^2/(k^2 + kk_c)$, it is not difficult to obtain

$$r_{e,h}(k) \approx -\frac{1 - \alpha_{e,h}^{1/2}(k/k_{co})^{1/2} - \alpha_{e,h}^{1} k/k_{co} + \alpha_{e,h}^{3/2}(k/k_{co})^{3/2}}{1 - \beta_{e,h}^{1/2}(k/k_{co})^{1/2} - \beta_{e,h}^{1} k/k_{co} + \beta_{e,h}^{3/2}(k/k_{co})^{3/2}},$$

$$\begin{aligned} r_{e,h}^2(k) &\approx 1 - \frac{2\xi(0,T)k_{co}(k/k_{co})^{1/2}/k_p}{1 - \beta_{e,h}^{1/2}(k/k_{co})^{1/2} - \beta_{e,h}^{1} k/k_{co} + \beta_{e,h}^{3/2}(k/k_{co})^{3/2}} \approx \\ &\approx 1 - \left(\beta_{e,h}^{1/2} - \alpha_{e,h}^{1/2}\right)(k/k_{co})^{1/2} - \left(\beta_{e,h}^{1/2} - \alpha_{e,h}^{1/2}\right)\beta_{e,h}^{1/2} k/k_{co} - \\ &- \left(\beta_{e,h}^{1/2} - \alpha_{e,h}^{1/2}\right)\left(\beta_{e,h}^{1} + \left(\beta_{e,h}^{1/2}\right)^2\right)(k/k_{co})^{3/2} + \left(\beta_{e,h}^{1/2} - \alpha_{e,h}^{1/2}\right)\left(\beta_{e,h}^{3/2} - 2\beta_{e,h}^{1/2}\beta_{e,h}^{1}\right)(k/k_{co})^2 + \ldots \end{aligned},$$

$$\alpha_e^{1/2} = \left(\frac{1}{p} - \xi(0,T)\right)\frac{k_{co}}{k_p},$$

$$\beta_e^{1/2} = \left(\frac{1}{p} + \xi(0,T)\right)\frac{k_{co}}{k_p},$$

$$\alpha_e^1 = \beta_e^1 = \frac{k_{co}}{k_p}\left(\frac{k_p}{k_{co}} - \frac{k_{co}}{k_p}\right),$$

$$\alpha_e^{3/2} = \beta_e^{3/2} = \frac{k_{co}}{2k_{co}^{1/2}k_p}\left(1 - \frac{k_{co}^2 p^2}{k_p^2}\right),$$

$$\alpha_h^{1/2} = (p - \xi(0,T))k_{cp} / k_p,$$

$$\beta_h^{1/2} = (p + \xi(0,T))k_{cp} / k_p,$$

$$\alpha_h^1 = \beta_h^1 = \frac{1}{2}\left(1 - \frac{k_{co}^2 p^2}{k_p^2}\right).$$

Here, we should use $p < p_{\max} = k_p / k_{co}$. Now we obtain an expansion of the type (44) in terms of half-integer and integer powers $(d/\delta_c)^\nu$, $\nu = -1/2, -1, -3/2....$, $\delta_c = 1/k_{co}$. Let us present the first term obtained during the approximation $r_{e,h}^{2m}(k) \approx 1 - 4m\xi(0,T)(kk_{co})^{1/2} / k_p$:

$$a_{1/2} = \left(2^{3/2}/9\right)\xi(0,T)\Gamma(4+1/2)\varsigma(3+1/2)k_{co} / \left(\varsigma(4)k_p\right).$$

The effect of graphene on a metal substrate in configuration *b* is reduced to the appearance of corrections $(k_\Omega d)^{-1}$, $(k_\Omega d)^{-2}$, …, which arise with their own coefficients in the parentheses of relations of the type (44).

## 5. Numerical results

The temperature dependencies of the Casimir pressure for two graphene sheets without a substrate are shown in Fig. 4. Formula (12) and the formulas for graphene conductivity given above, taking into account SD, were used (asterisks for curve 2). Curves 1,3,4 were obtained without considering SD. The influence of SD is weak. For curve 2, it is represented by the symbols ***. On average, the contribution of SD is less than 2% and is mainly manifested at small distances. SD slightly reduces the force. This can be explained by the fact that it makes a capacitive contribution to conductivity. At large distances, which correspond to long waves, the influence of SD is very small. It should be noted that dispersion interactions between graphene sheets and graphene with a number of structures have been studied in a number of works, for example, [36–41]. In some of them, for example, [36, 37], the comparison is made with respect to the asymptotic Casimir pressure. In this regard, it should be noted that the use of the temperature dependence $\xi(0,T)$ is strongly determined by the model of graphene conductivity. In paper [41], the equilibrium and nonequilibrium cases are considered when examining the polarization tensor for graphene within the framework of quantum field theory and the Dirac model of massless fermions in graphene. The analysis is based on the tensor dielectric permittivity.

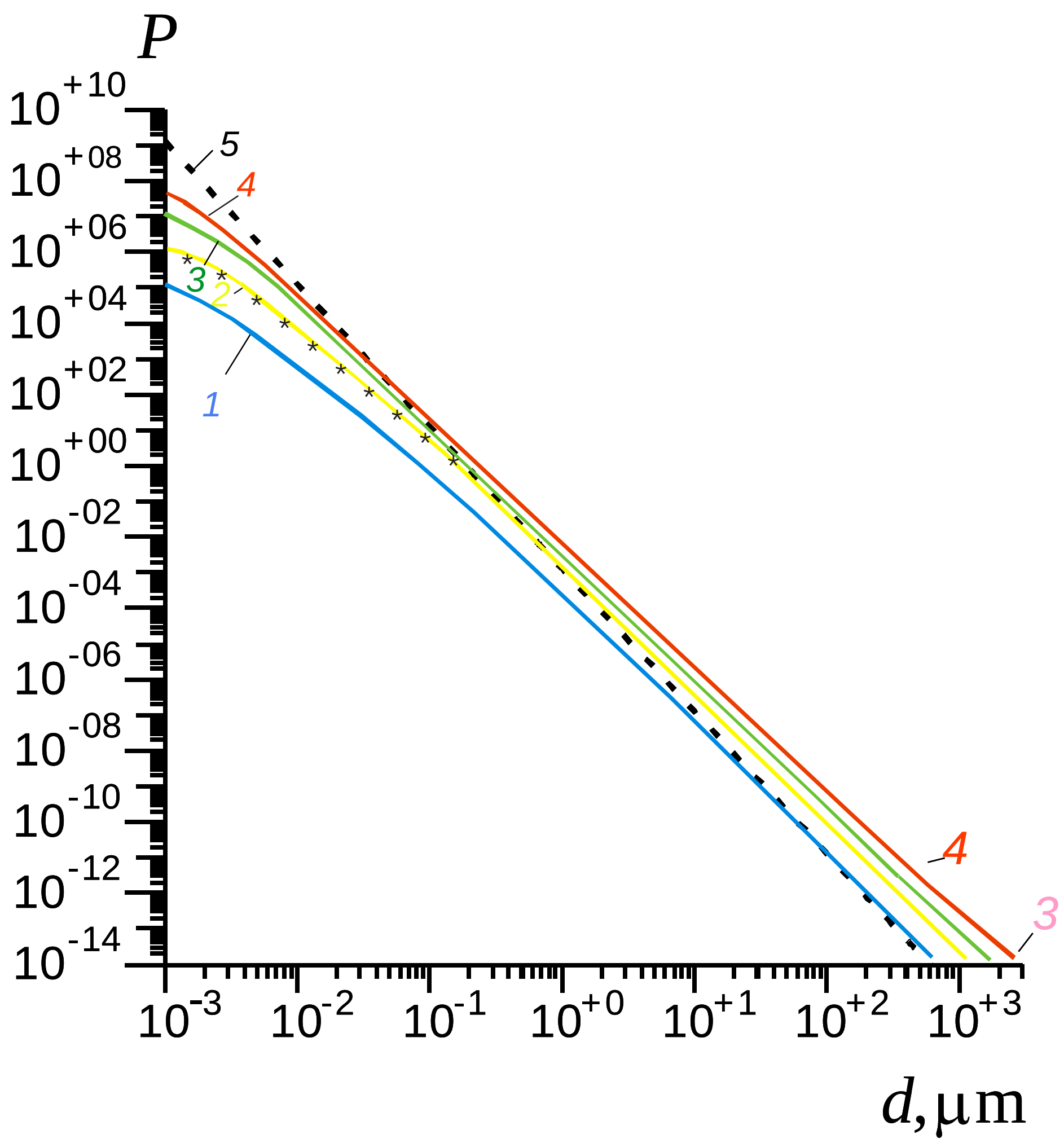

Fig. 4. Casimir pressure on two graphene sheets without a substrate at different temperatures: 100 K (curve 1), 300 K (2), 600 K (3), 900 K (4). The symbols *** show the influence of SD for curve 2. The curves are plotted using formula (12). 5 – the Casimir result

Fore $\xi(k,T)=\xi(0,T)/(1+k/k_c)=k_\Omega/(k+k_c)$ we can assess the strength in comparison with Casimir's result using the mean squares of the RCs. We have

$$r_e(k,\kappa)=-\frac{k_\Omega/(k+k_c)}{2k/\sqrt{k^2+\kappa^2}+k_\Omega/(k+k_c)},$$

$$r_h(k,\kappa)=-\frac{k_\Omega/(k+k_c)}{2\sqrt{k^2+\kappa^2}/k+k_\Omega/(k+k_c)}.$$

The average squares are estimated using the formula

$$\bar{r}_{e,h}^2 = \frac{1}{k_\Omega^2} \int_0^{k_\Omega} \int_0^{k_\Omega} r_{e,h}^2(k,\kappa) dk d\kappa .$$

Then the force is approximately in $\left(\bar{r}_e^2 + \bar{r}_h^2\right)/2$ times less than the Casimir force. A numerical estimate at room temperature and the corresponding carrier concentration yields a value of 0.0025.

**Funding of the work**

The study was carried out with financial support from the Ministry of Education and Science of Russia as part of the implementation of the state assignment (project No. FSRR-2026-0006). **Acknowledgments**

The author expresses gratitude to A.A. Petrunin for providing the results of DFT modeling.

**Conflict of interest**

The author reports the absence of a conflict of interest.

---